\documentclass[manuscript,screen]{acmart}

\AtBeginDocument{%
  }

\setcopyright{acmlicensed}
\copyrightyear{2018}
\acmYear{2018}
\acmDOI{XXXXXXX.XXXXXXX}

\acmISBN{978-1-4503-XXXX-X/18/06}

\usepackage{tabularray}
\usepackage{array}
\usepackage{pifont}
\usepackage{tcolorbox}
\usepackage{rotating}
\usepackage{enumitem}
\usepackage{makecell}
\usepackage{array}
\usepackage{graphicx}
\usepackage{multirow}
\usepackage{ragged2e}

\newcommand{\fixed}[1]{\textcolor{black}{#1}}
\newcommand{\hotfix}[1]{\textcolor{black}{#1}}

\newcommand{\eg}{\textit{e.g., }}
\newcommand{\ie}{\textit{i.e., }}

\begin{document}

\title[Detecting Refactoring Commits in Machine Learning Python Projects]{Detecting Refactoring Commits in Machine Learning Python Projects: A Machine Learning-Based Approach}


\author{Shayan Noei}
\orcid{0000-0002-5675-7817}
\email{s.noei@queensu.ca}
\affiliation{%
  \institution{Queen's University}
  \city{Kingston}
  \country{Canada}}

\author{Heng Li}
\email{heng.li@polymtl.ca}
\orcid{0000-0001-5441-6763}
\affiliation{%
  \institution{Polytechnique Montréal}
  \city{Montréal}
  \country{Canada}
}
\author{Ying Zou}
\email{ying.zou@queensu.ca}
\orcid{0000-0002-5335-0261}
\affiliation{%
  \institution{Queen's University}
  \city{Kingston}
  \country{Canada}}





\renewcommand{\shortauthors}{Noei et al.}

\begin{abstract}
\label{abstract}
Refactoring aims to improve the quality of software without altering its functional behaviors. Understanding developers' refactoring activities is essential to improve software maintainability. 
The use of machine learning (ML) libraries and frameworks in software systems has significantly increased in recent years, making the maximization of their maintainability crucial. 
Due to the data-driven nature of ML libraries and frameworks, they often undergo a different development process compared to traditional projects. As a result, they may experience various types of refactoring, such as those related to the data. \fixed{The state-of-the-art refactoring detection tools have not been tested in the ML technical domain, and they are not specifically designed to detect ML-specific refactoring types (\eg data manipulation) in ML projects;} therefore, they may not adequately find all potential refactoring operations, specifically the ML-specific refactoring operations.
Furthermore, a vast number of ML libraries and frameworks are written in Python, which has limited tooling support for refactoring detection. PyRef, a rule-based and state-of-the-art tool for Python refactoring detection, can identify 11 types of refactoring operations with relatively high precision. In contrast, for other languages such as Java, state-of-the-art tools are capable of detecting a much more comprehensive list of refactorings. For example, Rminer can detect 99 types of refactoring for Java projects.
Inspired by previous work that leverages commit messages to detect refactoring, we introduce MLRefScanner, a prototype tool that applies machine-learning techniques to detect refactoring commits in ML Python projects. MLRefScanner detects commits involving both ML-specific refactoring operations and additional refactoring operations beyond the scope of state-of-the-art refactoring detection tools.
To demonstrate the effectiveness of our approach, we evaluate MLRefScanner on 199 ML open-source libraries and frameworks and compare MLRefScanner against other refactoring detection tools for Python projects. Our findings show that MLRefScanner outperforms existing tools in detecting refactoring-related commits, achieving an overall precision of 94\% and recall of 82\% for identifying refactoring-related commits. MLRefScanner can identify commits with ML-specific and additional refactoring operations compared to state-of-the-art refactoring detection tools.
When combining MLRefScanner with PyRef, we can further increase the precision and recall to 95\% and 99\%, respectively.
MLRefScanner provides a valuable contribution to the Python ML community
, as it allows ML developers to detect refactoring-related commits more effectively in their ML Python projects. Our study sheds light on the promising direction of leveraging machine learning techniques to detect refactoring activities for other programming languages or technical domains where the commonly used rule-based refactoring detection approaches are not sufficient.

\end{abstract}



\maketitle

\section{introduction}
\label{introduction}


Understanding the appropriate timing and implementation of refactoring in software development enables developers to comprehend software evolution, make informative decisions, and enhance their knowledge of code design~\cite{atwi2021pyref, alomar2021refactoring}. However, it is challenging to gain insights into maintenance practices, as developers often provide insufficient documentation regarding their refactoring activities. Quite often, developers use general terms (\eg clean-up) to describe the refactoring actions instead of providing specific details~\cite{alomar2021refactoring}.

Sophisticated tools ({\it e.g., }Rminer~\cite{tsantalis2018accurate, tsantalis2020refactoringminer}) are available for other programming languages, such as Java. However, refactoring detection tools for Python are relatively limited. Moreover, existing refactoring detection tools are not specifically designed or tested to identify domain-specific (\eg ML) refactorings. The existing refactoring detection tools primarily rely on pattern matching using Abstract Syntax Tree (AST)~\cite{tsantalis2018accurate, tsantalis2020refactoringminer, atwi2021pyref} to detect refactoring operations. Consequently, they are unable to identify accessory information in the code, such as logical refactorings that do not follow a specific pattern. These refactorings are comprehended by developers but cannot be identified by pattern-matching in the AST structure changes.

Python has become one of the most popular programming languages, especially for developing Artificial Intelligence (AI) and Machine Learning (ML) libraries, frameworks, and applications~\cite{mihajlovic2020use, srinath2017python, TIOBE_2022, raschka2020machine}. The utilization of ML frameworks and libraries has increased in recent years, and many software projects now depend on ML libraries and frameworks~\cite{dilhara2021understanding, gevorkyan2019review}. Therefore, ensuring the proper maintainability of the ML libraries and frameworks is essential. Refactoring is a technique used to improve the maintainability of software~\cite{al2017empirical, arif2020refactoring}. Due to the data-driven nature of ML projects, they undergo a different development process, including design and implementation, which is more complex than traditional software development~\cite{wan2019does, amershi2019software, de2019understanding}. Various refactoring experiences and types may exist in ML-specific contexts, e.g., data modifications like switching dataset types (\eg from CSV to Parquet) or altering the way of iterating through dataset records.

Compared to other programming languages, Python prioritizes readability~\cite{Python.org, khoirom2020comparative} and has a simpler coding structure~\cite{abdulkareem2021evaluating, khoirom2020comparative}. For instance, Python is a dynamically typed language, while languages like Java are statically typed \cite{khoirom2020comparative}. Consequently, identifying class and object relationships through static code analysis in Python is affected. Abstract syntax tree in Python provides less detailed information compared to languages like Java. Moreover, there are no expectations to improve Python AST building tools in the near future~\cite{dilhara2022discovering}.

\fixed{Existing approaches for refactoring detection can be classified into two categories: rule-based and keyword-based.} The state-of-the-art keyword-based approaches~\cite{alomar2019can, ratzinger2007mining, di2018preliminary, kim2014empirical} involve searching through commit messages to identify refactoring commits based on a predefined set of keywords. For instance, if certain keywords, such as \textit{refactor}, appear in a commit message, it is identified as a refactoring commit. However, not all refactoring activities are explicitly mentioned in commit messages. Therefore, keyword-based approaches may fail to detect refactoring commits that lack explicit keyword mentions in their commit messages.

PyRef~\cite{atwi2021pyref} is the most effective and validated refactoring detection tool for Python. PyRef leverages pattern-matching within the Abstract Syntax Tree (AST) using a predefined set of rules~\cite{atwi2021pyref}. PyRef can detect up to 11 types of refactoring operations, covering 11\% of the refactoring operations that Rminer can detect in Java code. PyRef achieved a precision of 89.6\% and a recall of 76.1\%.  
Additionally, PyRef has not been specifically designed or validated for the ML technical domain for refactoring detection, which may provide incomplete coverage of refactoring operations and increase the likelihood of false negatives in identifying refactoring commits.

In this paper, we introduce MLRefScanner, a purpose-built prototype tool designed specifically to identify refactoring commits in the history of ML Python projects. Leveraging machine learning techniques and algorithms, MLRefScanner aims to improve the coverage of identifying refactoring operations in Python code and enhance the capabilities of detecting ML-specific refactorings. MLRefScanner allows practitioners and researchers to track and analyze refactoring activities with high precision and recall in the software development history of ML Python libraries and frameworks, enabling a deeper understanding of the scope and evolution of the ML codebase.

To evaluate the effectiveness of MLRefScanner, we conduct empirical studies on 199 ML libraries and frameworks. In this study, we aim to answer the following research questions:

\noindent\textbf{RQ1. What is the performance of our approach for identifying refactoring commits in ML Python projects?}--- To identify refactoring commits in the history of ML Python libraries and frameworks, 
we analyze their commit history and extract multiple dimensions of representative features ({\it e.g.,} commit message-related features) to train MLRefScanner with different classifiers ({\it e.g., }Random Forest). \fixed{Our findings demonstrate that LightGBM is the best-performing classifier for detecting refactoring commits, enabling MLRefScanner to achieve an overall precision of 94\%, a recall of 82\%, and an AUC of 89\%.  MLRefScanner can identify refactoring commits with ML-specific and additional refactoring types compared to state-of-the-art tools.}

\noindent\textbf{RQ2. What are the main features for explaining refactoring commits?}--- To gain deeper insights into the nature of refactoring commits for practitioners and increase the transparency of the proposed approach, we apply a features evaluation approach using the number of decision splits and the LIME interpreter~\cite{ribeiro2016should}, which helps us identify the most important features of refactoring commits. As a result, the top 10 features that indicate a refactoring commit include: (1) textual vocabulary features: {\it update, remove, refactor, improve, and move}; (2) process features: {\it author's refactoring contribution ratio, lines deleted, lines added, code entropy}; and (3) code features: {\it lines of declarative code}.

\noindent\textbf{RQ3. Can we leverage commit information to complement existing keyword-based and rule-based approaches}?--- We compare MLRefScanner with existing keyword and rule-based approaches in identifying refactoring commits, demonstrating its superiority. Our approach outperforms the existing best approach ({\it i.e., }PyRef) achieving a 22\% increase in AUC. Additionally, we explore the potential benefits of utilizing ensemble learning to complement PyRef with MLRefScanner. The ensemble learning approach achieves an overall precision of 95\% and a recall of 99\%, significantly improving the low recall of the state-of-the-art approaches.

Our work makes the following main contributions: 
\begin{itemize}
    \item We propose a learning-based approach and a prototype, MLRefScanner, for detecting refactoring commits in ML Python projects, achieving an overall precision of 94\% and recall of 82\%.
    \item We propose an ensemble approach that combines our learning-based approach and a state-of-the-art rule-based refactoring detection approach ({\it i.e., } PyRef), further improving the performance of both MLRefScanner and PyRef.
    \item We provide a carefully curated dataset that labels refactoring commits in ML Python projects.
    \item We address the low recall issue of the existing state-of-the-art approaches by considering different types of features ({\it i.e., }textual, process, and code) in refactoring commit detection.
    \item We present the first prototype of an ML-specific refactoring commit detection tool, \fixed{capable of identifying commits related to ML-specific refactoring types as well as additional commits involving general (not ML-specific) refactoring types.}
\end{itemize}

\noindent 
The replication package of the study can be accessed at: \url{https://github.com/seal-replication-packages/TOSEM2024}


\noindent \textbf{Organization.} The remainder of our study is organized as follows.
Section~\ref{experiment_setup} describes the experiment setup of this study. Section~\ref{results} presents the motivation, approaches, and results of our research questions. Section~\ref{sec:discussions} explains the generalizability of our approach to other technical domains.
Section~\ref{sec:implications} provides the implications of our study.
Section~\ref{sec:threads_to_validity} discusses the threats to the validity of our findings.  Section~\ref{sec:related_work}
surveys related studies and compares them to our work.
Finally, we conclude our paper and present future research directions in Section~\ref{sec:conclusion}.

\section{experiment setup}
\label{experiment_setup}
This section presents the setup of our study, including our data collection and data analysis approaches.
\begin{figure}
    \includegraphics[width=\textwidth]{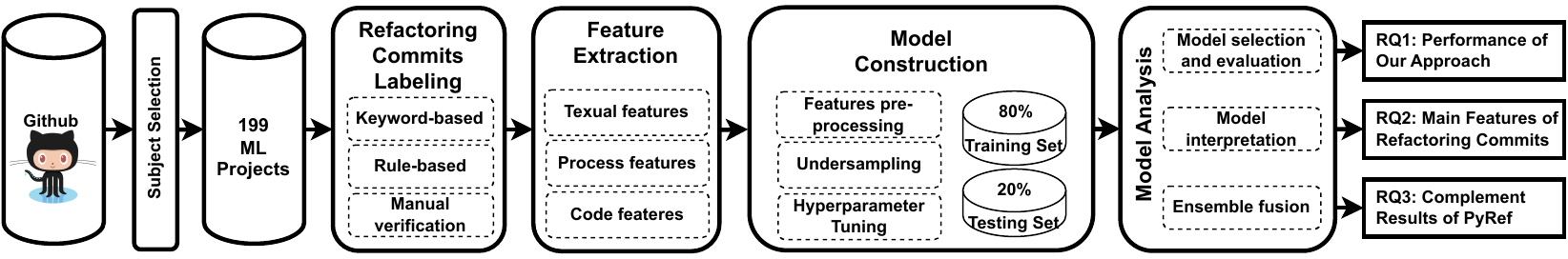}
    \caption{Overview of the approach}
    \label{fig:overview}
\end{figure}
\subsection{Overview of Our Approach}
We propose an ML-based approach to detect refactoring commits in ML Python projects. An overview of our study is depicted in Figure \ref{fig:overview}. 
We train MLRefScanner in the machine-learning domain using 199 ML libraries and frameworks. Leveraging the commit history on the subject projects, we create a curated dataset by labeling each commit as a refactoring commit or not, based on the commit messages and refactoring analysis results obtained from the PyRef tool~\cite{atwi2021pyref}. We manually validate our labeling approach through a validation process involving two individuals to assess its performance. Subsequently, we extract textual, process, and code features to capture the characteristics of each commit. We then partition the data into training sets and testing sets. During the training phase, we preprocess the data by applying undersampling to address issues related to an unbalanced dataset. Additionally, we conduct hyperparameter tuning to optimize the training settings of our classifiers. We interpret the trained model and extract the most important features to provide deeper insights into the characteristics of refactoring and non-refactoring commits. Finally, we ensemble the proposed model with the state-of-the-art refactoring detection tool ({\it i.e., } PyRef) and assess whether we can enhance the refactoring detection performance by considering a variety of features in each commit.

\subsection{Subject Selection}
\label{sec:subject_selection}
We select machine learning projects to train and test MLRefScanner in the ML technical domain. Inspired by previous work~\cite{openja2022studying}, we use the advanced search on GitHub\footnote{\url{https://docs.github.com/en/rest/search?apiVersion=2022-11-28}} to select repositories of machine learning libraries, frameworks, and applications primarily written in Python by specifying keywords, including {\it deep learning, reinforcement learning,} and {\it machine learning}. As a result, we identify 161,662 projects. It is time-consuming to extract refactoring commits. For example, it takes PyRef\footnote{We use PyRef in our data curation, as described in Section~\ref{sec:labeling}.} approximately one day to process each project because it involves parsing the commit code change history and constructing an AST for each commit. We plot the distributions of project size (lines of code), stars, and forks of the repositories and adopt a systematic process~\cite{openja2022studying} to select the projects. The selection criteria help us select mature and active projects that have demonstrated sustained growth and community engagement. We remove the projects that have a size smaller than 23,148 KB, fewer than 27 stars, or fewer than 18 forks as such projects are below the 3rd quartile of the distribution; have had no pushes in the year leading up to the data collection date (from 2021/04/01 to 2022/04/01); have no issues; have no downloads; are archived; are disabled; or have fewer than 1,000 commits. As a result, we obtain 217 repositories. Furthermore, we perform a manual validation to remove 18 irrelevant (\ie non-ML repositories or repositories that contain only documentation). Finally, 199 repositories are selected as the basis ({\it i.e.,} training and testing) for our experimental analysis.

\subsection{Labeling Refactoring Commits}
\label{sec:labeling}
To label commits as either refactoring or non-refactoring in the selected ML projects, we employ two state-of-the-art automated approaches: (1) keyword-based labeling and (2) rule-based labeling with PyRef. If a commit is identified as a refactoring by either approach, we label it as a refactoring commit. During the labeling process, we classify 155,780 commits as refactoring and 343,187 as non-refactoring commits. Then, we perform manual validation on \fixed{a statistically significant sample size} to verify the precision and recall of the automated labeling results.

\subsubsection{Keyword-based labeling} We compile a list of keywords that indicate a refactoring commit ({\it e.g.,  modify, enhance,} or {\it clean-up}) used in previous studies~\cite{alomar2019can, di2018preliminary, stroggylos2007refactoring, ratzinger2007mining}.
These keywords are typically associated with self-admitted refactoring operations, meaning that developers explicitly mention in the commit messages that refactoring is performed.
To create a starting point, we manually investigate each keyword using a small set of 10 sample commits that contain the keyword and assess if they consistently represent refactoring commits. This validation process helps us minimize false positive keywords and the manual validation cost. We exclude keywords that do not consistently relate to refactoring operations. For instance, we find that certain keywords, such as {\it fix} might not always represent refactoring operations. The final set of chosen keywords used to label a refactoring commit includes:  {\it move, refactor, remove, rename, split, clean, improve, unused, cleanup, simplify, restruct, inline, parameterize, consolidate, encapsulate,} and {\it update}. We then use the keywords to match the commit messages of all the commits of the studied projects: a commit is labeled as a refactoring commit if it contains at least one of the keywords. We use an exact match search, which sets a commit as a refactoring commit if any part of the commit message contains the specified keywords.
In the labeling process, we also classify commits that do not alter any executable code ({\it e.g.,} only changes in the ReadMe file) as non-refactoring commits to ensure that the labeled refactoring commits involve changes to the executable code.

\begin{table}
\centering
\footnotesize
\caption{
The set of features utilized in the study. Textual features include vocabulary and features extracted from commit messages; process features encompass details processed from commits; and code features comprise code metrics extracted from the source code before and after each commit code changes. (Detailed description of each feature is included in the replication package)}
\label{tbl:studied-metrics}
\begin{tabular}{>{\hspace{0pt}}m{0.15\linewidth}>{\hspace{0pt}}m{0.787\linewidth}}
\textbf{Feature Type} & \textbf{Name of Features}                                                                                                                                                                                                                                                                                                                                                                                                                                                                                                                                                                                                                              \\ 
\hline
(1) Textual              & TF Vectorized Commit Messages (Vocabulary), Words Count, Sentences Count, Readability                                                                                                                                                                                                                                                                                                                                                                                                                                                                                                                                                     \\ 
\hline
(2) Process              & Lines Added, Lines Deleted, Lines Changed, Number of Files, Code Entropy, Refactoring Contribution Ratio, Has Executable File                                                                                                                                                                                                                                                                                                                                                                                                                                                                                                                         \\ 
\hline
(3) Code                 & Avg Count Line, Avg Count Line Blank, Avg Count Line Code, Avg Count Line Comment, Avg Cyclomatic, Count Class Base, Count Class Coupled, Count Class Coupled Modified, Count Class Derived, Count Decl Class, Count Decl Executable-Unit, Count Decl File, Count Decl Function, Count Decl Instance Method, Count Decl Instance Variable, Count Decl Method, Count Decl Method All, Count Line, Count Line Blank, Count Line Code, Count Line Code Decl, Count Line Code Exe, Count Line Comment, Count Stmt, Count Stmt Decl, Count Stmt Exe, Cyclomatic, Max Cyclomatic, Max Inheritance Tree, Max Nesting, Ratio Comment To Code, Sum Cyclomatic 
\end{tabular}
\\[-10pt]
\end{table}

\subsubsection{Rule-based labeling} 
Not all refactoring operations are self-admitted; some may be performed without explicit mention in the commit message. As PyRef can achieve a precision of 89.6\% and a recall of 76.1\% in detecting refactoring operations in ML Python projects, we utilize PyRef \cite{atwi2021pyref} to account for refactoring activities not mentioned in the commit message. If PyRef identifies any refactoring operations, we label the commit as a refactoring commit even if there are no specific refactoring keywords in the commit message. The results obtained from PyRef complement the keyword-based labeling results. As PyRef can detect 11 types of refactoring operations including {\it rename method, add parameter, remove parameter, change/rename parameter, extract method, inline method, move method, pull up method,} and {\it push down method}. PyRef is used to label our dataset in identifying commits that involve these refactoring operations.

\subsubsection{Manual validation}
\label{sec:validating_labeling}
To assess the accuracy and reliability of our labeling approach, we engage one author and a third-year undergraduate computer science student to conduct manual validation. Before splitting our dataset into training and testing sets, we randomly select 383 commits from the dataset to ensure that they are not included in either the testing or training sets.
This selection is made to maintain a 95\% confidence level with a 5\% margin of error. Both validators independently assess the commit messages and code changes of the commits and label them based on whether they observe a refactoring operation or not. If they observe one, they label it as a refactoring commit. The Cohen's kappa agreement level~\cite{cohen_statistical_1988} is found to be 0.64 after the initial labels, indicating a moderate level of agreement~\cite{mchugh2012interrater} between the validators. Both validators discuss the initial labels they provided on each commit to reach a consensus on a final label. The final manually validated set includes 75 refactoring commits and 308 non-refactoring commits. By comparing the labels of 383 commits agreed upon by the validators and the labels generated by our automatic labeling approach for the same 383 commits, we achieve a precision of 93\% and a recall of 93\% using our automatic labeling approach (\textit{i.e.,} using the agreed-upon manual labeling results as the ground truth). In detail, the automated labeling process identifies 5 false-positive and 5 false-negative labels. We observe that the main reason for the false negatives is that the PyRef is incapable of identifying all refactoring types, thus failing to detect refactoring operations in certain commits and reporting them as non-refactoring. Moreover, the reason for the false positives is the misinterpretation of certain keywords, such as \textit{improve}, which do not necessarily describe refactoring commits. For instance, we observe a commit with the message ``fix(helper): improve collision in random\_port'' containing the \textit{improve} keyword, but it is not related to refactoring. Consequently, we use our manually labeled results as the ground truth for further analysis and evaluation of our approach.

\subsection{Feature Extraction}
We consider three dimensions of features to identify refactoring commits, including (1) textual features, (2) process features, and (3) code features, as listed in Table~\ref{tbl:studied-metrics}.

\subsubsection{Textual features}
\label{vectorize_commit_messages}
The textual features are listed in Table~\ref{tbl:studied-metrics}. We first clean up the commit messages to identify the vocabulary in the commit messages. We convert all messages to lowercase to ensure that the capitalization of words does not affect the classification. To address typos and mistakes in commit messages, we utilize the Enchant library~\cite{Thomas} to perform spelling corrections and unify word usage, regardless of whether they are mistakenly written or not. We remove URLs, HTML tags, punctuations, emojis, numerical digits from the messages, and stop words ({\it e.g.,} an, with, or but). Moreover, we stem the commit sentences to their root~\cite{balakrishnan2014stemming} form to ensure that variations of words ({\it e.g.,} fixing and fix) are treated as the same. We choose stemming to avoid duplication and improve the generalization of the features. For instance, when using Lemmatization, {\it refactor, refactoring,} and {\it refactored} are considered separate features. In contrast, with stemming, we handle all of the variations of the word with a single vector as {\it refactor}, which prevents our model from being overwhelmed by the explosion of the contextual meanings of unique words in the commits. Term frequency (TF)~\cite{sparck1972statistical} in commit messages is used to retain the ability to interpret and understand the contributions of individual words in the classification process, which assigns a value of 0 or 1 based on the presence or absence of a word. We opt for TF instead of TF-IDF (term frequency-inverse document frequency) because TF-IDF tends to assign lower weights to the most repeated words. However, in our case, certain words like {\it refactor} may appear frequently and should be weighted more. Additionally, we choose not to use vector transformation algorithms like Bert~\cite{devlin2018bert} or word2vec~\cite{mikolov2013linguistic} as they are not reversible and interpretable. Using such algorithms would prevent us from investigating the root cause and important features that distinguish a commit based on whether it involves refactoring or not. In the vectorization setup, we use 6-grams (\ie a maximum of 6 consecutive words as one term) to extract sufficient information from the commit messages, thereby improving our ability to identify potential refactoring activities. The choice of 6-grams is based on a previous study on self-admitted refactorings~\cite{alomar2019can}, which identify the maximum number of words in a sentence that can potentially reveal a refactoring commit as 6. In addition, we calculate textual features, such as word count and sentence count for each commit message. To assess the readability of these commit messages, we employ the Flesch reading ease metric~\cite{flesch1948new}, which indicates the level of comprehension difficulty in the commit messages.

\subsubsection{Process features}
\label{sec:process_metrics}
We collect a set of process features that capture various characteristics of commits, as outlined in Table~\ref{tbl:studied-metrics}. We analyze the modified code blocks and files within each commit using the \textit{git log} information on the cloned repositories to extract features including {\it lines added, lines deleted, lines changed, has executable file,} and {\it number of files}. This information helps us better describe the scope of code changes in each commit. The refactoring contribution ratio (RCR) provides information on the refactoring history and experience of each developer. To calculate the RCR, we collect \hotfix{all labeled commits of a developer before each commit} (described in Section~\ref{sec:labeling}) and utilize the formula below:

\begin{equation}
    \begin{aligned}
        RCR(\textit{i}) =\frac{\textit{Refactoring commits of the author (i)}}{\textit{All commits of the author (i)}}
    \end{aligned}
\end{equation}

\noindent To gain insights into the complexity and structural changes within each commit, we calculate the code changes entropy using Shannon's entropy~\cite{shannon1948mathematical}, which has been utilized for measuring the complexity of code changes~\cite{hamou2008measuring}. For each file within a commit with executable code, we calculate the relative code churn of that file by dividing the code churn of that file by the total code churn of the commit, denoted as P($x_{i}$). Then, we use Shannon's entropy formula to calculate the code change entropy within each commit. The formula for Shannon's entropy is as follows:
\begin{equation}
    \begin{aligned}
        H(x)= -\sum_{i}^{}P(x_{i})logP(x_{i})
    \end{aligned}
\end{equation}

\subsubsection{Code Features}
To capture information about code changes within each commit, such as code complexity, we calculate code features related to the code changes of each commit. We extract the modified files associated with each commit. Subsequently, we calculate 32 sets of code features in class, function, and code levels, as outlined in Table~\ref{tbl:studied-metrics}, before and after each commit. We use the Understand\footnote{\url{https://scitools.com/}} tool to calculate code features~\cite{medeiros2023trustworthiness, zhou2021investigating, vatanapakorn2022python} for the current and previous state of each commit, then we compare the current and previous state of changes and calculate the difference of the code features ({\it i.e., } reduction or increase) for each code feature.

\subsubsection{Feature pre-processing}
\label{sec:pre_processing}
The existence of highly correlated and redundant features could affect the stability of the model and the interpretation of features~\cite{noei2019too, harrell2001regression}. Therefore, on the training set (described in Section~\ref{sec:data_spliting}), we conduct correlation analysis and redundancy analysis of the process and code features. We don't perform correlation and redundancy analysis on the textual features as it may lead to some loss of information and nuances during model interpretation. Each textual feature can provide critical insights into why the model has reached a particular decision, and removing correlated textual features can present challenges when interpreting the decision-making process of the model.
\begin{itemize}[leftmargin=10pt]
\item\noindent\textbf{\textit{Correlation analysis:}}
        Our features do not follow a normal distribution, therefore we utilize Spearman's correlation coefficient to determine the correlation between the computed features. A coefficient $>0.7$ indicates a strong correlation~\cite{noei2019too, nguyen2010studying}. For strongly correlated features, we choose to retain one feature in our model while removing the other. The results of the correlation analysis show that {\it Number of Files, CountClassBase, CountDeclExecutableUnit, CountLine, CountLineCode, CountLineCodeExe, CountStmt, CountStmtDecl,} and {\it CountStmtExe} are highly correlated with other features. Therefore, we exclude them from our features. The dendrogram of our correlations analysis is included in our replication package.
    
\item\noindent \textbf{\textit{Redundancy analysis:}}
    R-squared is a metric used to indicate how much of the variance in a dependent variable can be explained by independent variables~\cite{miles2005r}. To identify redundant features that can be estimated from other features, we use one feature as a dependant and other features as independent features~\cite{pintas2021feature} setting a cut-off value of 0.9~\cite{jiarpakdee2016study} for the R-squared. Following the removal of highly correlated features, we conduct the redundancy analysis, yet no redundant features are found. 

\item\noindent \textbf{\textit{Removing low-variance features:}}
We exclude features with a variance threshold lower than 0.001 in the training set. This threshold implies that the feature value is identical in 99.9\% of the data, indicating a minor impact on the data. This step becomes necessary because our initial textual features include 3,433,690 features. By excluding low-variance features, we reduce the feature size to 651, thus focusing on informative features.
\end{itemize}
Finally, we use min-max normalization~\cite{hazewinkel2001minimax} on the process and code features to fit them into the same range as TF vectorized textual features, ensuring that the features are rescaled to the range [0, 1]. For instance, the number of file changes within each commit could vary from [0, $\infty$]. By utilizing min-max normalization, we rescale it to the range [0, 1].

\subsection{Model Construction}
\label{sec:model_construction}
To build our ML-based approach, we discuss the model construction steps in the following subsections.

\subsubsection{Data splitting}
\label{sec:data_spliting}
We randomly split our dataset into a 20\% testing set and an 80\% training set. To avoid bias, we leave the testing set untouched throughout the entire model-building process, including features pre-processing, undersampling, and hyperparameter tuning.

\subsubsection{Under-sampling of training data}
\label{sec:under_sampling}
Our dataset contains a total of 155,780 refactoring commits and 343,187 non-refactoring commits, resulting in an imbalanced dataset. Class imbalance in our dataset has the potential to skew our classification model towards the majority class ({\it i.e.,} non-refactoring commits), leading to reduced accuracy \cite{kotsiantis2006handling}. As the size of the dataset is large ({\it i.e., }498,975 records with 651 features), oversampling would significantly increase the computational costs. Moreover, undersampling has been shown to be a better approach compared to oversampling~\cite{drummond2003c4, mishra2017handling}. Hence, we choose to undersample our training dataset. Random undersampling can lead to data loss during the random selection of majority class samples~\cite{batista2004study}. Therefore, we perform three different NearMiss undersampling approaches~\cite{mani2003knn} and evaluate our models on each, seeking the best-performed undersampling approach. We apply the following undersampling approach to our dataset:
\begin{itemize}[leftmargin=10pt]
    \item NearMiss-1: Choose negative (\ie non-refactoring) samples with an average \fixed{Euclidean distance}~\cite{mani2003knn, dokmanic2015euclidean} closest to the sum of the three closest positive (\ie refactoring) samples.
    \item NearMiss-2: Select negative samples that are nearest to the sum of all closest positive examples.
    \item NearMiss-3: For each positive sample, select the negative examples that are closest.
\end{itemize}

\subsubsection{Model Training}
\label{sec:used_models}
To train our model, we use eleven different classifiers belonging to three different categories. (1) Linear classifiers: Naive Bayes, Support Vector Machine (SVM); (2) Nonlinear classifiers: Decision Tree, k-Nearest Neighbor, Random Forest, Multi-Layer Perceptron (MLP), and CatBoost; and (3) Ensemble Learning (EL): Adaptive Boosting (AdaBoost), Gradient Boosting Machine (GBM), XGBoost, and LightGBM~\cite{moran2022important, jafarzadeh2021bagging}.

\subsubsection{Hyperparameter Tuning} 
To enhance the accuracy of the selected models, we perform hyperparameter tuning on the training set. Our dataset consists of approximately 499,293 records, containing the extracted features listed in Table~\ref{tbl:studied-metrics}. Training the models can be time-consuming, with an average training time of around 20 minutes per model. Therefore, we utilize random search to explore the hyperparameter space and find the best configuration efficiently. Random search, an efficient and effective hyperparameter tuning approach, allows us to sample combinations efficiently and often achieves better results in less time compared to other approaches, such as grid search~\cite{bergstra2012random}.
\section{Results}
\label{results}
In this section, we provide the motivation, approach, and findings
for each of our research questions.
\subsection{RQ1. What is the performance of our approach for identifying refactoring commits in ML Python projects?}
\label{results:RQ1}
\subsubsection{Motivation}
Refactoring is applied in practices to increase software maintainability. It is essential to track previous refactoring activities by analyzing commit information so that developers can gain a better understanding of the state of code. This includes monitoring the history and timing of refactoring applied in the history of code, identifying vulnerable code segments that have undergone frequent refactoring operations, and estimating future maintenance needs~\cite{cedrim2017understanding, nyamawe2020feature, armijo2022refactoring}.
The state-of-the-art refactoring detection tools for Python code are relatively limited. For instance, PyRef can detect 11 types of refactoring in Python code. Moreover, existing tools utilize AST pattern matching to detect refactoring in general repositories, however, the existing tools have not been specifically designed and tested for particular domains, such as machine learning.
With the rapid growth in the number and usage of ML libraries and frameworks, predominantly written in Python, our approach, MLRefScanner is developed to detect commits with more comprehensive refactoring operations within the ML libraries and frameworks written in Python.


\subsubsection{Approach} 
\label{sec:RQ1_approach}
\fixed{As explained in Section~\ref{experiment_setup}, we train and evaluate our refactoring commit detection models using our dataset, which contains 199 ML projects.}
We assess the overall performance of our model based on precision, recall, AUC, and F1 scores. Precision defines the ratio of true positives to the total positive predictions. Recall measures the ratio of true positives to the actual positive instances. F1 score calculates the harmonic mean of precision and recall. AUC represents the ability of the model to distinguish between classes based on probability~\cite{hand2012assessing}. 

\noindent\textbf{Model Evaluation.}
We evaluate the performance of our model in the following scenarios:
\begin{itemize}[leftmargin=10pt]
    \item \textbf{Mixed Projects:} 
    \fixed{We train our refactoring commit detection models using 80\% of the commits of the selected projects, and then we assess the performance of our model on the testing set which consists of the remaining 20\% of all commits in our dataset from different projects altogether.}
    \item \textbf{Cross Projects:} \fixed{Quite often, some projects may not have a sufficient number of historical refactoring commits to train a model. Therefore, we train our model using all but one (n-1 projects) and assess the performance of our models in the remaining project as testing, where the model is tested on the project excluded from the training dataset. We repeat the leaving-one-out approach for all projects.}
    \item \textbf{Within Projects:} 
    \fixed{To maintain consistent results across different projects and prevent large projects from dominating our dataset, we employ a within-projects setting and test our models separately for each project. We train our model using 80\% of the commits randomly sampled from each project and evaluate its performance on the remaining 20\% of commit messages from the same project. Specifically, we train one model for each project and test it on the same project.}
    \item \textbf{Ground Truth:} \fixed{To ensure the reliability of our model's performance, we test the final model, which is trained on 80\% of all commits from all projects, on the ground truth dataset as described in Section~\ref{sec:validating_labeling}.}
\end{itemize}

\noindent\textbf{Manual Analysis of the Identified Refactoring Types in the mixed-projects Setting.}
To gain deeper insights into the capability of MLRefScanner in detecting ML-specific refactoring operations, employing a 95\% confidence level with a 5\% margin of error, we select a set of 383 commits identified as refactoring by MLRefScanner. Then, we ask one of the co-authors and a fourth-year undergraduate computer science student to conduct a thematic analysis to inspect the refactoring patterns identified by the tool in the chosen refactoring commits. Subsequently, they engage in discussions to compare the patterns they have identified and agree on final refactoring patterns in each commit. During the manual validation, the validators consider a portion of code as a refactoring if it does not alter the external behavior of the code~\cite{fowler1999refactoring}. \fixed{In manual thematic analysis, Cohen's kappa agreement~\cite{cohen_statistical_1988} level in commits involving ML-specific or general refactoring activities is determined to be 0.82 between validators, indicating a perfect level of agreement~\cite{mchugh2012interrater}.}

\subsubsection{Findings} 

\begin{figure}
    \includegraphics[width=0.5\linewidth]{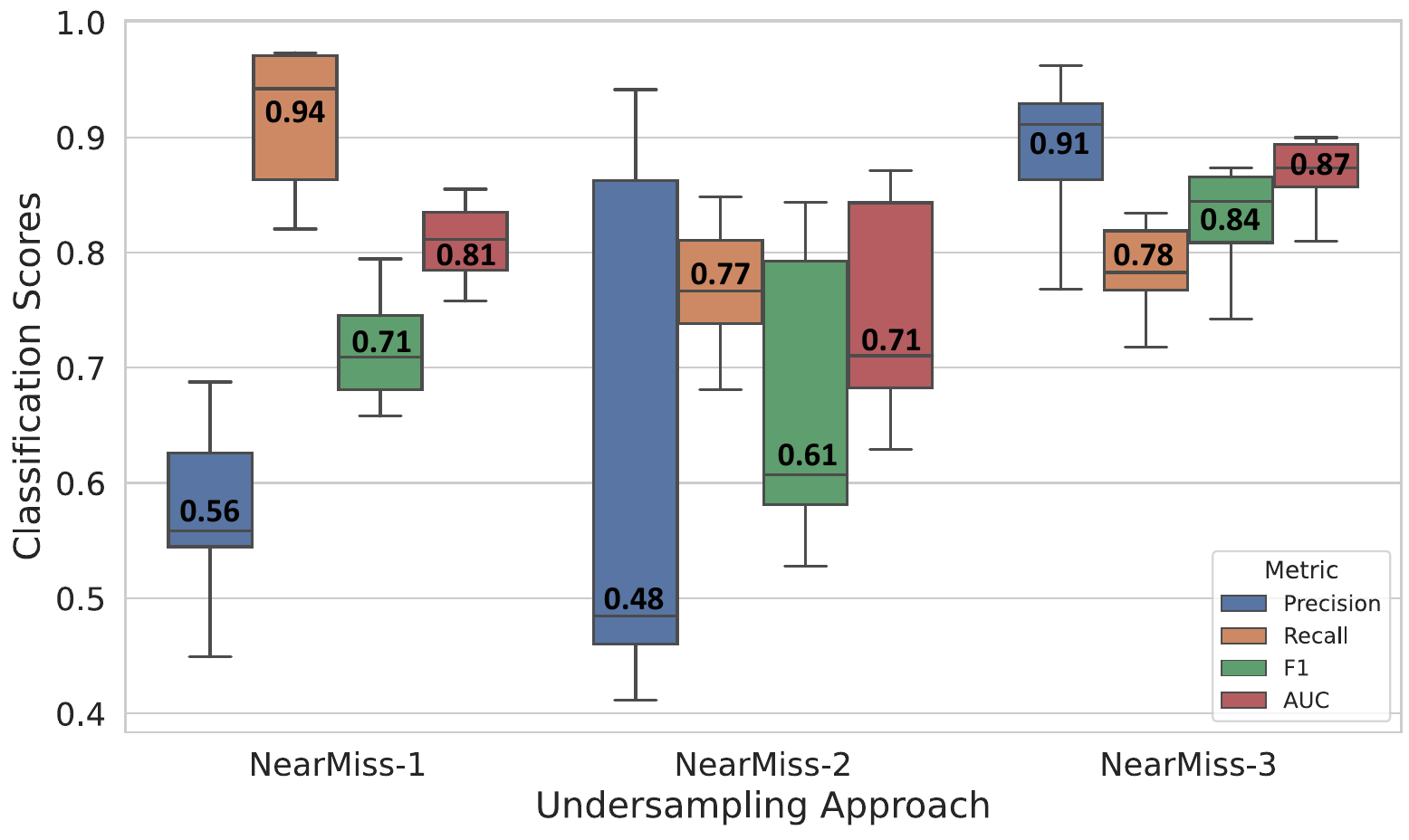}
    \caption{Compare the Results of Different Undersampling Approaches}
    \label{fig:compare-undersampling}
\end{figure}

\noindent\textbf{Best Undersampling Approach: } \textbf{ \textit{NearMiss-3} undersampling provides the highest AUC (87\%) and F1 (0.84\%) scores in the mixed-projects setting.} Figure~\ref{fig:compare-undersampling} presents the evaluation scores for 11 different classifiers, {\it NearMiss-3} undersampling achieves the highest median precision (91\%) and a median recall of 78\%. On the other hand, {\it NearMiss-1} has a higher median recall (94\%), but it experiences a significantly lower precision (56\%). Additionally, we manually inspect the classification scores in all the classification models (details in the replication package) and verify the superiority of {\it NearMiss-3} for all classifiers on our dataset. This can be attributed to the complexity of the decision boundary, as NearMiss-3 selects non-refactoring samples that are closest to refactoring samples, a technique that has also been demonstrated as effective in prior studies~\cite{ghimire2021machine, mani2003knn, fonseca2021addressing}. Consequently, we choose {\it NearMiss-3} as the undersampling technique.

\begin{table}
\centering
\footnotesize
\caption{The evaluation scores of different classifiers.}
\label{tbl:compare-scores}
\begin{tabular}{>{\hspace{0pt}}m{0.3\linewidth}>{\hspace{0pt}}m{0.16\linewidth}>{\hspace{0pt}}m{0.12\linewidth}>{\hspace{0pt}}m{0.1\linewidth}>{\hspace{0pt}}m{0.1\linewidth}}
\textbf{Model}         & \textbf{Precision} & \textbf{Recall} & \textbf{AUC} & \textbf{F1}  \\ 
\hline
\textbf{LightGBM}      & 0.94               & 0.82            & 0.89         & 0.87         \\ 
\hline
\textbf{CatBoost}      & 0.92               & 0.83            & 0.9          & 0.87         \\ 
\hline
\textbf{GBMClassifier} & 0.90                & 0.83            & 0.90          & 0.87         \\ 
\hline
\textbf{AdaBoost}      & 0.91               & 0.82            & 0.89         & 0.86         \\ 
\hline
\textbf{XGBoost}       & 0.96               & 0.76            & 0.88         & 0.85         \\ 
\hline
\textbf{SVM}           & 0.93               & 0.77            & 0.87         & 0.84         \\ 
\hline
\textbf{MLPClassifier} & 0.93               & 0.77            & 0.87         & 0.84         \\ 
\hline
\textbf{Random Forest} & 0.84               & 0.78            & 0.86         & 0.81         \\ 
\hline
\textbf{K-Neighbors}   & 0.88               & 0.74            & 0.85         & 0.81         \\ 
\hline
\textbf{Decision Tree}          & 0.79               & 0.81            & 0.86         & 0.80          \\ 
\hline
\textbf{Complement NB} & 0.77               & 0.72            & 0.81         & 0.74        
\end{tabular}
\\[-10pt]
\end{table}
\noindent\textbf{Best Classification Model: } 
\textbf{The LightGBM classifier exhibits the best performance in refactoring commit identification, achieving an overall precision of 94\%, recall of 82\%, and AUC of 89\% on the mixed-projects setting.}
Table~\ref{tbl:compare-scores} presents the details of the scores obtained for different classifiers using NearMiss-3 undersampling. Our findings demonstrate that both LightGBM and CatBoost classifiers outperform other approaches, with CatBoost achieving a precision of 92\% and a recall of 83\% in the mixed-projects setting. Given our objective to maximize precision (\textit{i.e., }minimize false positives), we select LightGBM as the nominated model and the best-performing classifier for the remainder of our study.  

\begin{figure}
    \includegraphics[width=0.8\linewidth]{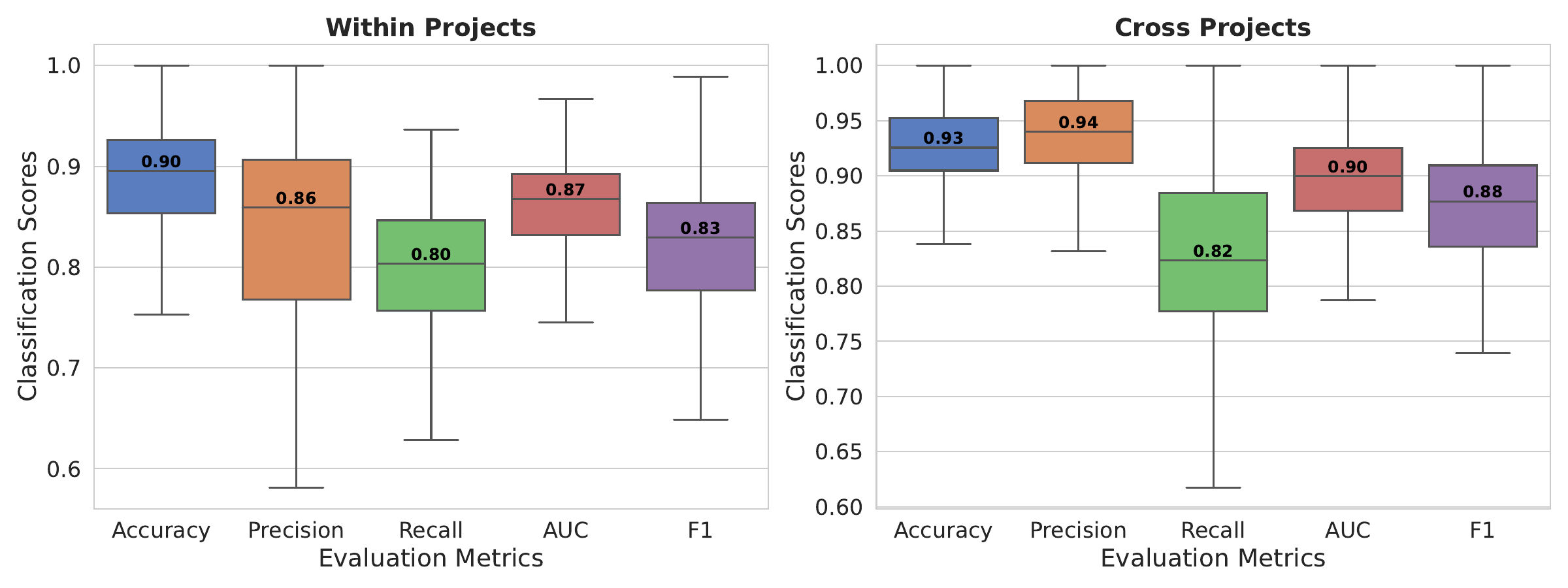}
    \caption{\hotfix{Compare the Performance of Our Approach Using Cross and within-projects Validations.}}
    \label{fig:validation_general}
\end{figure}

\noindent\textbf{Our approach can not only accurately detect refactoring commits when trained and tested in the same projects (\textit{i.e.,} within-projects), but also in projects that have not been seen before, even if they may exhibit different characteristics (\textit{i.e.,} cross-projects).} In within-projects validation, we achieve a median precision of 94\%, a median recall of 82\%, and a median AUC of 87\%. In the cross-projects validation, we achieve a median precision of 86\%, a median recall of 80\%, and a median AUC of 90\%. The results of the within-projects and cross-projects validations are shown in Figure~\ref{fig:validation_general}. Furthermore, we achieve a precision of 93\%, a recall of 93\%, and an AUC of 96\% on the ground truth dataset.

\fixed{\noindent\textbf{MLRefScanner can detect commits with ML-specific or general refactoring operations that cannot be detected by state-of-the-art refactoring detection tools.} Through manual thematic analysis, we identify a set of ML-specific and general refactoring patterns in the commits detected by MLRefScanner that cannot be identified using state-of-the-art AST pattern-matching refactoring detection tools. ML-specific refactoring patterns are typically related to dataset handling, data reading/writing operations, mathematical computations, and ML model parameter adjustments. On the other hand, general refactorings are broader refactoring operations that do not adhere to a regular pattern; instead, they mainly rely on domain knowledge.
For instance, code optimization involves replacing a block of code with a simpler and more understandable alternative. The explanation and the frequency of application for the extended refactoring types are provided in Table~\ref{tbl:ml-specific}.}

\fixed{\noindent\textbf{MLRefScanner is trained on Python projects but is not limited to Python source files and code.} Through manual thematic analysis, we observe that 22 out of 384 commits with source files other than Python have been detected as refactoring commits. This is because MLRefScanner is trained not only using code metrics but also a set of process and textual features that are independent of programming languages.}

\begin{table}
\centering
\footnotesize
\caption{\fixed{The list of refactoring types that can be identified by MLRefScanner but cannot be identified by state-of-the-art pattern-matching detection tools. The examples of each refactoring type are included in our replication package.}}
\label{tbl:ml-specific}
\begin{tabular}{l|lll}
\multicolumn{1}{l}{} & \textbf{Type} & \textbf{Description} & \textbf{Frequency} \\ 
\hline
\multirow{7}{*}{\textbf{ML-Specific}} & Data Handling  Optimization & Optimizing data reading/writing from/to the dataset. & 14.4\% \\ 
\cline{2-4}
 & Model Initialization Refinement & Modify the hyper-parameter initializations in the models. & 04.8\% \\ 
\cline{2-4}
 & Resource Allocation Optimization & Adjusting hardware usage parameters. & 04.8\% \\ 
\cline{2-4}
 & Data Path Management & Managing dataset/plots/models storage paths. & 03.8\% \\ 
\cline{2-4}
 & Data Type Clarification & Converting data types from one to another (e.g., from NumPy to DataFrame). & 02.9\% \\ 
\cline{2-4}
 & Data Presentation Enhancement & Refactoring data visualization code. & 02.9\% \\ 
\cline{2-4}
 & Mathematical Operation Refactoring & Changing mathematical calculation to a simpler form. & 00.7\% \\ 
\hline
\multirow{7}{*}{\textbf{General}} & Code Cleanup & Removing unused code or files. & 16.7\% \\ 
\cline{2-4}
 & Code Simplifications & Replacing code with more simplified alternatives. & 13.5\% \\ 
\cline{2-4}
 & Import/Export Optimization & Removing or relocating unused dependencies. & 10.9\% \\ 
\cline{2-4}
 & Logging Enhancement & Improving log messages for user understanding. & 10.5\% \\ 
\cline{2-4}
 & Requirement/Configuration Update & Updating program requirements or configurations. & 08.4\% \\ 
\cline{2-4}
 & Condition Simplification & Simplifying complex conditional statements. & 03.0\% \\ 
\cline{2-4}
 & File/Dependency Path Refinement & Optimizing/updating dependency/file path. & 02.5\% \\

\end{tabular}
\end{table}

\begin{center}
\begin{tcolorbox}[colback=lightgray!10, colframe=lightgray, arc=4pt, left=2pt, right=2pt, top=2pt, bottom=2pt]
MLRefScanner can accurately detect refactoring commits within the ML technical domain (\ie ML projects) in mixed-projects, within-projects, cross-projects, and ground truth settings. MLRefScanner achieves 94\% precision, 82\% recall, and 89\% AUC on the testing set of the selected ML projects in the mixed-projects setting. \fixed{MLRefScanner is capable of detecting ML-specific (\textit{e.g.,} data processing optimization) and a set of new refactoring types (\textit{e.g.,} Code Cleanup) that are dismissed by state-of-the-art tools relying on AST pattern matching algorithms.}
\end{tcolorbox}
\end{center}

\subsection{RQ2. What are the main features for explaining the refactoring commits?}
\label{results:RQ3}
\subsubsection{Motivation}

In the first research question, we propose a classifier capable of identifying refactoring commits in the ML technical domain. It is important to explain the outcomes from the classifier decisions on detecting refactoring commits. By understanding key features, we can provide practitioners, especially tool makers, and researchers, with insights into the characteristics of refactoring commits.

\subsubsection{Approach}

\textbf{Importance of each feature type.} We assess the impact of each feature type (\textit{i.e.,} textual, process, and code) listed in Table~\ref{tbl:studied-metrics} on the decision-making process of our best-performing model (\textit{i.e.,} LightGBM). To achieve this, we train separate LightGBM models using textual, process, and code data individually on the ML training set and measure the performance of each model on the ML testing set separately. This enables us to (1) indicate the importance of each feature type and (2) demonstrate how each feature type contributes to the identification of refactoring commits.

\textbf{Individual feature importance.} We select the best-performing trained classifier (\ie LightGBM) to compute the importance of each feature by assessing its frequency and significance in splitting the decision trees during the training process. However, the number of splits in the decision tree doesn't interpret how each feature contributes more towards predicting specific prediction ({\it i.e., } refactoring or non-refactoring commit) and the model remains a black box. Therefore, we leverage the local interpretable model-agnostic explanations (LIME) technique~\cite{ribeiro2016should}. 

LIME enables us to explain the features of individual instances in our dataset and assigns weights to each feature, indicating their impact on the decision-making process of the model. Using LIME, we gain insights into the model's behavior at the instance level and can better understand how each feature affects the predictions of our model. This approach enhances the transparency and interpretability of the decision-making process of our model. 

\textbf{Explaining feature contributions with LIME.} We first use the classifier's splitting score, and then we apply LIME to extract the weight of each feature for each instance in our testing set. A negative weight reported for a feature indicates that the presence or increase in that feature contributes to the decision-making process toward the negative class \textit{(i.e.,} refactoring), while a positive weight indicates that the increase or presence of that feature is associated with the decision-making process favoring the toward class (\textit{i.e.,} non-refactoring). Lastly, a neutral weight implies that the feature does not independently influence the decision-making process but rather relies on other features for classification.
We train LIME on our training set and then apply it to every instance in the testing set. We use the median weights of features, and we assign a positive, negative, or neutral weight to each feature, indicating the extent to which a feature is generally associated with refactoring commits.

\subsubsection{Findings}

 \textbf{Overall, textual features play the most important role in identifying refactoring commits. However, the addition of process and code features to textual features results in an improvement of 7\% in precision, 11\% in recall, and 7\% in AUC.} 
 \hotfix{The results of inspecting the impact of different types of features ({\it i.e.,} process, code, and textual) on our model are shown in Table~\ref{tbl:compare-metrics}. The significant improvement in recall brought by the process and code features enables the identification of refactoring commits that may not be explicitly mentioned in the commit messages.}

\begin{table}
\centering
\footnotesize
\caption{The results of the trained classifier on each metric type and overall}
\label{tbl:compare-metrics}
\begin{tabular}{>{\hspace{0pt}}m{0.2\linewidth}>{\hspace{0pt}}m{0.18\linewidth}>{\hspace{0pt}}m{0.134\linewidth}>{\hspace{0pt}}m{0.117\linewidth}>{\hspace{0pt}}m{0.05\linewidth}}
\textbf{Feature Type} & \textbf{Precision} & \textbf{Recall} & \textbf{AUC} & \textbf{F1}  \\ 
\hline
\textbf{Textual}   & 0.87               & 0.71            & 0.83         & 0.78         \\ 
\hline
\textbf{Process}      & 0.52               & 0.73            & 0.71         & 0.60         \\ 
\hline
\textbf{Code}         & 0.53               & 0.41            & 0.62         & 0.46         \\ 
\hline
\textbf{Overall}      & 0.94               & 0.82            & 0.89         & 0.87        
\end{tabular}
\\[-10pt]
\end{table}

\hotfix{\textbf{The appearance of certain vocabulary keywords in a commit message indicates the existence of refactoring in that commit.} The most influential keywords for classifying a commit as refactoring include {\it update, remove, refactor, improve, move, rename, clean, cleanup, add, simplify, split, unused, restructure, option, name, parameter,} and {\it argument}. Specifically, some vocabulary features may indicate a general refactoring operation in a commit (\textit{e.g., refactor}), while others may be more specific, such as \textit{rename}, which indicates the renaming of a variable, method, or class at some level of the code.}

\begin{table*}
\centering
\caption{The table presents the most important features in our classifier, each with at least 10 splits in the decision tree. Features marked with \ding{52} indicate refactoring features, \ding{56} indicate non-refactoring features --- signifies neutral features, and features with the CMV prefix indicate vocabulary features. For the vocabulary features, we provide an example of a commit message containing that feature and set the possible refactoring operation in parentheses. For other features, we include a description of the feature.}
\label{tbl:important_features}
\footnotesize
\resizebox{\linewidth}{!}{%
\begin{tabular}{l|llll}
\multicolumn{1}{l}{\textbf{Feature Type}} & \textbf{Feature Name} & \textbf{Importance} & \textbf{Refactoring} & \textbf{Example / Description} \\ 
\hline
\multirow{23}{*}{\textbf{Textual }} & \textbf{CMV:updat (update)} & 170 & \ding{52} & The return type of the method call is updated (Change Return Type) \\ 
\cline{2-5}
 & \textbf{CMV: remov (remove)} & 145 & \ding{52} & The parameters of the method is/are removed (Remove Parameter) \\ 
\cline{2-5}
 & \textbf{CMV:refactor} & 78 & \ding{52} & Refactored the contributor guide into many more sub sections and made a start on some of them \\ 
\cline{2-5}
 & \textbf{CMV:improv} & 50 & \ding{52} & Small name update to improve clarity in func\_wrapper (Change Function Name) \\ 
\cline{2-5}
 & \textbf{CMV:move} & 45 & \ding{52} & Move dtype functions from general to dtype jax (move method) \\ 
\cline{2-5}
 & \textbf{CMV:renam (rename)} & 43 & \ding{52} & The method is\_array is renamed to is\_native\_array (Rename Method) \\ 
\cline{2-5}
 & \textbf{Readability} & 41 & — & The Flesch commit message readability metric~\cite{flesch1948new} (described in Section~\ref{sec:process_metrics}) \\ 
\cline{2-5}
 & \textbf{Words Count} & 41 & — & Number of words in the commit message \\ 
\cline{2-5}
 & \textbf{CMV:clean} & 38 & \ding{52} & Clean up databricks scripts (Multiple Refactoring Operations) \\ 
\cline{2-5}
 & \textbf{CMV:cleanup} & 32 & \ding{52} & Fixed typos in code cleanup process (Change Variable Name) \\ 
\cline{2-5}
 & \textbf{CMV:add} & 29 & \ding{52} & The parameters are added to the method get\_relevant\_items\_by\_threshold (Pull Up Parameters) \\ 
\cline{2-5}
 & \textbf{CMV:simplifi (simplify)} & 23 & \ding{52} & Simplification of predict function through parameter (Remove Parameter) \\ 
\cline{2-5}
 & \textbf{CMV:split} & 22 & \ding{52} & Split configuration and modeling files (Move Method) \\ 
\cline{2-5}
 & \textbf{CMV:unus (unused)} & 19 & \ding{52} & Remove unused args parameter from PreTrainedConfig.from\_pretrained (Remove Parameter) \\ 
\cline{2-5}
 & \textbf{CMV:readme} & 15 & \ding{56} & Rename the readme file \\ 
\cline{2-5}
 & \textbf{CMV:restructur (restructure)} & 14 & \ding{52} & Restructured function (Move Method) \\ 
\cline{2-5}
 & \textbf{CMV:option} & 14 & \ding{52} & Add options for save\_as (Add Parameter) \\ 
\cline{2-5}
 & \textbf{CMV:doc} & 12 & \ding{56} & Update document of standalone deployment \\ 
\cline{2-5}
 & \textbf{CMV:name} & 12 & \ding{52} & Rename some function (Rename Method) \\ 
\cline{2-5}
 & \textbf{CMV:ad} & 11 & — & Import optional modules ad hoc in the runner \\ 
\cline{2-5}
 & \textbf{CMV:paramet (parameter)} & 10 & \ding{52} & Change parameter names (Rename Parameter) \\ 
\cline{2-5}
 & \textbf{CMV:fix} & 10 & \ding{56} & Fix parallel execute for bug \\ 
\cline{2-5}
 & \textbf{CMV:argument} & 10 & \ding{52} & Removed unused argument (Remove parameter) \\ 
\hline
\multirow{6}{*}{\textbf{Precess}} & \textbf{Refactoring Contribution} & 239 & \ding{52} & The developer refactoring contribution (described in Section~\ref{sec:process_metrics}) \\ 
\cline{2-5}
 & \textbf{Lines Deleted} & 140 & \ding{52} & The number of lines deleted in a commit \\ 
\cline{2-5}
 & \textbf{Has Executable File} & 108 & — & If a commit contains executable code \\ 
\cline{2-5}
 & \textbf{Lines Added} & 94 & \ding{52} & The number of lines added in the commit \\ 
\cline{2-5}
 & \textbf{Code Entropy} & 88 & \ding{52} & The code entropy of each commit (described in Section~\ref{sec:process_metrics}) \\ 
\cline{2-5}
 & \textbf{Lines Changed} & 49 & — & The number of lines changed in a commit (Lines Added - Lines Deleted) \\ 
\hline
\multirow{8}{*}{\textbf{Code}} & \textbf{CountLineCodeDecl} & 87 & \ding{52} & Count of lines that consist of declarative source code~\cite{Understand} \\ 
\cline{2-5}
 & \textbf{CountDeclFunction} & 51 & — & Methods declared within a class accessible exclusively via an instance of that class~\cite{Understand} \\ 
\cline{2-5}
 & \textbf{SumCyclomatic} & 42 & \ding{52} & The total weight of methods belonging to a class~\cite{Understand} \\ 
\cline{2-5}
 & \textbf{AvgCountLineCode} & 36 & \ding{52} & Mean number of lines containing source code within all nested functions or methods~\cite{Understand} \\ 
\cline{2-5}
 & \textbf{CountLineComment} & 30 & \ding{52} & Mean number of lines containing comments for all nested functions or methods~\cite{Understand} \\ 
\cline{2-5}
 & \textbf{CountLineBlank} & 24 & \ding{52} & The number of black lines of code~\cite{Understand} \\ 
\cline{2-5}
 & \textbf{RatioCommentToCode} & 22 & — & Proportion of lines containing comments compared to lines containing code~\cite{Understand} \\ 
\cline{2-5}
 & \textbf{MaxCyclomatic} & 19 & \ding{52} & The highest cyclomatic complexity among all nested functions or methods~\cite{Understand}
\end{tabular}
}
\\[-10pt]
\end{table*}

\hotfix{\textbf{Past refactoring contributions of the developers, code complexity features, and code changes are the most common indicators of refactoring.} More specifically, increased values of the process and code features such as  \textit{refactoring contribution, lines deleted, lines added, code entropy, CountLineCodeDecl, SumCyclomatic, AvgCountLineCode, MaxCyclomatic, CountLineComment, } and \textit{CountLineBlank} play a significant role in influencing the decision of our model towards classifying a commit as refactoring.}

\textbf{The presence of certain vocabulary keywords in a commit message indicates the absence of refactoring in that commit.} Keywords such as \textit{readme, doc,} and \textit{fix} are identified as characteristics that lead the model to make a non-refactoring decision. Furthermore, other features, including \textit{Executable File}, \textit{CountDeclFunction},\textit{ Lines Changed}, \textit{Readability}, \textit{words count}, \textit{RatioCommentToCode}, and {\it ad} vocabulary keyword, are considered dependent features that assist other features in the decision-making process. They do not directly influence the model's decision but contribute to the overall classification outcome in conjunction with other features. \fixed{They enhance the model's predictions by adding additional context that complements other features, thereby playing a supportive role.}
The results of our feature importance approach with at least 10 splits in the decision tree are listed in Table~\ref{tbl:important_features} (the full list of features is included in our replication package). 

Our approach provides a complementary set of important textual, process, and code features, helping practitioners, researchers, and tool makers understand the significance of different features in identifying refactoring commits. 

\begin{center}
\begin{tcolorbox}[colback=lightgray!10, colframe=lightgray, arc=4pt, left=2pt, right=2pt, top=2pt, bottom=2pt]
Textual features, which capture the characteristics of commit messages, play a crucial role in detecting refactoring commits. Additionally, past refactoring contributions of developers, code changes, and code complexity features significantly contribute to identifying refactoring commits that lack indicators in their commit messages.

\end{tcolorbox}
\end{center}

\subsection{RQ3. Can we leverage commit information to complement existing keyword-based and rule-based approaches?}
\label{results:RQ2}
\subsubsection{Motivation}
Historical information of software has been used to identify refactorings in prior studies~\cite{alomar2019can, ratzinger2007mining, di2018preliminary, kim2014empirical, atwi2021pyref}. Some approaches use a specific set of keywords to determine whether a commit is related to refactoring or not~\cite{alomar2019can, ratzinger2007mining, di2018preliminary, kim2014empirical}. Moreover, certain tools, like PyRef~\cite{atwi2021pyref}, utilize pattern matching on the AST level to ascertain the presence of refactoring in a commit. In this research question, we first compare our approach with state-of-the-art tools and methods that identify refactoring commits to evaluate the effectiveness of our approach. Moreover, we aim to determine whether ensembling our approach with PyRef, the state-of-the-art and rule-based approach, aids in improving refactoring commit detection. This analysis helps us provide a more effective solution by complementing the state-of-the-art tools to identify refactoring commits in ML projects mainly written in Python.

\subsubsection{Approach}
\textbf{Comparing MLRefScanner with baselines.} \fixed{We compare the output of our approach with the following keyword-based and rule-based approaches by measuring the performance metrics ({\it i.e., } precision, recall, F1, and AUC) on the testing set in the mixed-projects setting.}

For the keyword-based approaches~\cite{alomar2019can, ratzinger2007mining, di2018preliminary, kim2014empirical}, we search through the entire commit history and identify refactoring commits based on the keywords provided. We include the set of keywords that each paper uses to identify refactoring commits in the replication package. 

For the rule-based approach~\cite {atwi2021pyref}, we run PyRef accordingly to get refactoring commits. 
As discussed in Section~\ref{sec:rq2_finding}, PyRef achieves the highest precision (100\%) compared to its opponents, but it suffers from low recall. This motivates us to consider ensembling our approach with PyRef to potentially improve the low recall of the PyRef.

\hotfix{\textbf{Ensembling MLRefScanner with PyRef.}} In our approach, we obtain a best-performing classifier that utilizes code, textual, and process features to identify whether a commit is a refactoring commit or not. Conversely, PyRef employs pre-defined rules to identify different types of refactoring and, consequently, refactoring commits. To potentially improve the recall in predicting refactoring commits, we propose an ensemble learning~\cite{rokach2010ensemble} method to combine the results from the best-performing classifier (\textit{i.e.,} MLRefScanner) with PyRef and measure if we can complement PyRef and improve its low recall. Through this ensemble analysis, both our classifier and PyRef vote on whether a commit is refactoring or not. We test and apply unanimous ({\it i.e., }both PyRef and our approach detect refactoring) and majority ({\it i.e., }at least one of PyRef or our approach detects refactoring) voting schemes and measure the enhancements in refactoring commit detection. This provides a more robust and comprehensive decision-making process.

\subsubsection{Findings}
\label{sec:rq2_finding}
\textbf{MLRefScanner outperforms the state-of-the-art rule-based and keyword-based approaches, achieving, on average, 36\% higher precision, 52\% higher recall, 28\% higher AUC, and 30\% higher F1 scores compared to previous keyword-based approaches~\cite{alomar2019can, ratzinger2007mining, di2018preliminary, kim2014empirical} and 6\% lower precision, 49\% higher recall, and 22\% higher AUC, compared to the rule-based state-of-the-art approach~\cite{atwi2021pyref} on the testing set.}
Table~\ref{tbl:compare-other-approaches_testing} shows the results of our approach compared to other refactoring detection approaches on our ML testing set in the mixed-projects setting.
MLRefScanner achieves a recall of 82\%, surpassing PyRef~\cite{atwi2021pyref} at 33\%, AlOmar \textit{et al.}\cite{alomar2019can} at 39\%, Ratzinger \textit{et al.}\cite{ratzinger2007mining} at 26\%, Zhang \textit{et al.}\cite{di2018preliminary} at 29\%, and Kim \textit{et al.}\cite{kim2014empirical} at 25\%. \hotfix{This indicates that previous approaches overlook a significant number of refactoring commits, making MLRefScanner superior to state-of-the-art methods in detecting all possible refactoring commits.}

\textbf{Our approach can complement the low recall of PyRef with a 60\% increase in its recall in the majority vote setting.} Despite a 6\% lower precision, our approach achieves a 49\% higher recall compared to PyRef, indicating our approach can capture a broader window of refactoring commits in the development history.
Table~\ref{tbl:ensemble} shows our ensembling results by combining our approach and PyRef. Our ensemble approach achieves a precision of 95\%, recall of 99\%, an AUC of 98\%, and an F1 score of 97\% in the majority vote setting. Therefore, we can effectively complement and assist PyRef in finding the refactoring commits that might have otherwise been overlooked.

\begin{center}
\begin{tcolorbox}[colback=lightgray!10, colframe=lightgray, arc=4pt, left=2pt, right=2pt, top=2pt, bottom=2pt]
On average, our approach, MLRefScanner, provides 21\% higher precision, 52\% higher recall, 28\% higher AUC, and 45\% higher F1 score compared to state-of-the-art rule-based and keyword-based approaches. MLRefScanner can complement PyRef's low recall and significantly increase it by 60\% in refactoring commit detection, motivating refactoring tool makers to consider more features in the refactoring detection process.
\end{tcolorbox}
\end{center}

\begin{table}
\centering
\footnotesize
\caption{Comparison of results obtained from MLRefScanner and state-of-the-art approaches.}
\label{tbl:compare-other-approaches_testing}
\begin{tabular}{>{\hspace{0pt}}m{0.203\linewidth}>{\hspace{0pt}}m{0.203\linewidth}>{\hspace{0pt}}m{0.162\linewidth}>{\hspace{0pt}}m{0.11\linewidth}>{\hspace{0pt}}m{0.079\linewidth}>{\hspace{0pt}}m{0.077\linewidth}}
\textbf{Appraoch} & \textbf{Category} & \textbf{Precision} & \textbf{Recall} & \textbf{AUC} & \textbf{F1} \\ 
\hline
\textbf{MLRefScanner} & ML Based & 94\% & 82\% & 89\% & 87\% \\ 
\hline
\textbf{PyRef~} & Rule Based & 100\% & 33\% & 67\% & 50\% \\ 
\hline
\textbf{AlOmar \textit{et al.}~} & Keyword Based & 39\% & 39\% & 56\% & 39\% \\ 
\hline
\textbf{Ratzinger \textit{et al.}~} & Keyword Based & 84\% & 26\% & 62\% & 40\% \\ 
\hline
\textbf{Zhang \textit{et al.} } & Keyword Based & 48\% & 29\% & 57\% & 36\% \\ 
\hline
\textbf{Kim \textit{et al.}~} & Keyword Based & 93\% & 25\% & 62\% & 39\%
\end{tabular}
\end{table}

\begin{table}
\centering
\footnotesize
\caption{Comparison of results obtained from Our classifier, PyRef, and ensemble approaches.}
\label{tbl:ensemble}
\begin{tabular}{>{\hspace{0pt}}m{0.31\linewidth}>{\hspace{0pt}}m{0.17\linewidth}>{\hspace{0pt}}m{0.12\linewidth}>{\hspace{0pt}}m{0.098\linewidth}>{\hspace{0pt}}m{0.08\linewidth}}
\textbf{Method}           & \textbf{Precision} & \textbf{Recall} & \textbf{AUC} & \textbf{F1}  \\ 
\hline
\textbf{PyRef~\cite{atwi2021pyref}}            & 100\%              & 39\%            & 70\%         & 56\%         \\ 
\hline
\textbf{MLRefScanner}      & 94\%               & 82\%            & 89\%         & 87\%         \\ 
\hline
\textbf{Unanimous Voting} & 100\%              & 21\%            & 61\%         & 35\%         \\ 
\hline
\textbf{Majority Voting}  & 95\%               & 99\%            & 98\%         & 97\%        
\end{tabular}
\\[-10pt]
\end{table}

\section{Discussion on the generalizability of our approach}
\label{sec:discussions}

\fixed{As discussed in Section~\ref{results:RQ1}, MLRefScanner performs well within ML projects and achieves high precision and recall under different testing settings in the ML technical domain compared to rule-based and keyword-based state-of-the-art approaches.} In this section, we want to investigate the potential applicability of MLRefScanner across diverse technical domains by assessing its performance on general (\ie non-ML) Python projects. For this purpose, we assemble a testing dataset comprising a variety of general Python projects, aiming to evaluate the adaptability of MLRefScanner beyond machine learning applications. We follow similar project selection criteria used for choosing machine learning projects to select the general Python projects (described in Section~\ref{sec:subject_selection}). We do not restrict the number of commits for general projects, allowing us to evaluate the effectiveness of our approach on both smaller projects with a limited number of commits and larger projects. This process results in a set of 12,665 general projects. Subsequently, we randomly select the same number of projects as in the machine learning category ({\it i.e.,} 199 projects) to evaluate the performance of our approach in general Python projects. We manually validate the general Python projects to ensure that projects containing only documentation are not included. Following this, we validate and verify our model to further assess the generalizability in the general Python projects. We assess the performance of our model in the context of general Python projects under the following scenarios:
\begin{itemize}[leftmargin=10pt]
\item \textbf{Combined General Projects:} Using a testing dataset comprising 199 general Python projects, we assess the performance of our model trained on ML projects across the entire dataset of general projects. This testing dataset includes all commits from all 199 general projects.
\item \textbf{Individual General Projects:} Using individual general Python projects as testing datasets, we individually test our model trained using ML projects, on all commits of each general Python project. This approach allows us to assess the effectiveness of our model within the specific context of each project.
\end{itemize}

\noindent\textbf{MLRefScanner can be extended to projects within other technical domains.} In the combined-general-projects setting, our model successfully achieves a precision of 99\%, a recall of 93\%, and an AUC of 96\% on general Python projects. In the individual-general-projects setting, we attain a median precision of 99\%, a median recall of 100\%, and a median AUC of 100\%. Maintaining high performance across technical domains is attributed to our use of features extractable from any Python repository, which are not specific to any particular domain. The performance of our approach can generalize from one technical domain (\textit{i.e.}, ML projects) to other domains (\textit{i.e.,} general Python projects). Practitioners and researchers can apply our findings in various contexts, whether they are working on traditional Python projects or those involving machine learning while maintaining a high level of performance.

\section{Implications}
\label{sec:implications}
In this section, we provide the implications of our study for managers, developers, and tool-makers.\\

\noindent\textbf{Enhancing refactoring evolution monitoring in the ML technical domain.} Accurately classifying refactoring commits is crucial for monitoring the activities in the software evolution and providing enhanced tracking mechanisms for stakeholders. \textbf{ML Software managers and developers} can effectively track the refactoring effort and gain insights into the volume of refactoring activities. This information allows them to review refactoring efforts and make well-informed decisions. For instance, they can identify areas of the code that are consistently being refactored, indicating potential issues that require prompt attention and action. Furthermore, developers' refactoring efforts can be observed and assessed, increasing motivation for refactoring. Existing tools and approaches for identifying refactoring activities suffer from low recall, either due to their limited ability to detect various types of refactoring or their primary reliance on text search-oriented methods. However, our approach facilitates effective tracking of refactoring commits in the history of ML Python projects with high precision and recall. Therefore, we encourage software teams to utilize our approach to track possible refactoring activities.

\noindent\textbf{Assisting ML researchers and refactoring tool makers in filtering out a substantial portion of commits that are unrelated to refactoring.} By drawing a clear distinction between refactoring and non-refactoring commits, our approach enables ML \textbf{researchers and developers} interested in refactoring to focus their efforts on commits that involve refactoring. We recommend researchers and developers refer to this targeted approach for designing automated tools and conducting in-depth studies on refactoring activities.

\noindent\textbf{Providing a set of features significantly correlated with refactoring activities.} By offering a set of features that exhibit a significant correlation with refactoring activities, we highlight the key attributes that encompass textual, process, and code feature aspects and warrant priority in the study of refactoring activities. \textbf{Toolmakers and researchers} can incorporate these features while studying refactoring evolution and related research. By presenting specific code features that correlate with refactoring activities, we aim to help \textbf{developers} focus on the critical areas that can be improved and refined to enhance the overall code quality and maintainability of the project.

\noindent\textbf{Demonstrating a complementary role of other features in refactoring identification.} The well-known refactoring detection tools PyRef~\cite{atwi2021pyref} in Python and Rminer~\cite{tsantalis2020refactoringminer} in Java utilize AST pattern matching algorithms. In this work, we showcase the effectiveness of incorporating source code features alongside pattern matching to enhance the accuracy and broaden the scope of refactoring types they can identify, we show how ensembling their rule-based approach with our ML-based approach can significantly increase the recall in refactoring commit detection. Therefore, we recommend \textbf{toolmakers} consider employing more hybrid techniques when designing refactoring identification tools for analyzing project development histories. Such hybrid approaches can yield more comprehensive and accurate results in detecting and understanding refactoring activities.
\section{Threats to Validity}
In this section, we discuss the threats to the validity of our study.\\

\label{sec:threads_to_validity}
\noindent\textbf{Threats to Construct Validity } 
concern the data preparation, feature selection, and model construction methods employed in our study. In dataset preparation, we utilize a set of textual, process, and code features to characterize the commit messages. Nevertheless, incorporating additional features has the potential to further enhance the accuracy of our classification model. In the model construction phase, we utilize random search for hyperparameter tuning in the model construction due to its cost-effectiveness, other hyperparameter tuning approaches could potentially further improve the accuracy of our classifier. \hotfix{To assess the performance of the models across commits with varying numbers of refactoring operations, we conduct a sensitivity analysis by adjusting the labeling threshold of the number of refactoring operations in a commit obtained from PyRef. 
We measure the performance of the models with different thresholds to evaluate the performance of the models in identifying refactoring commits. 
As the threshold for refactoring operations increases, the recall of the model drops, meaning that the classifier dismisses the commits with fewer numbers of refactoring operations. This observation further supports the efficacy of the selected threshold (\textit{i.e.,} the existence of at least one refactoring operation in a commit) in identifying refactoring commits.}

\noindent\textbf{Threats to Internal Validity } 
indicate the validity of methods utilized in our study. 
In the experiment setup, we opt to use term frequency (TF) to vectorize commit messages~\cite{luhn1957statistical} instead of using other vectorizing approaches, such as Bert~\cite{devlin2018bert} and word2vec~\cite{mikolov2013linguistic}, due to their lack of transparency and irreversibility. While we acknowledge that using such vector transformation algorithms could potentially enhance our commit message analysis, we prioritize the interpretability and reversibility of the chosen approach in order to provide explainability of our approach. 
In the process of labeling validation and ground truth preparation, we obtain a Cohen's kappa agreement level of 0.64, indicating a moderate level of agreement between the validators. However, the two validators have discussed the results of the manual validations meticulously to reach a conclusion on each commit and achieve a consensus on the final label, ensuring the avoidance of bias in the labeling process.

\noindent\textbf{Threats to External Validity } 
indicate the generalizability of our approach.
The vocabulary features that our classification model is trained on are tailored to English languages. The reported performance scores reflect the performance of our classification model on repositories that predominantly use English to document their commits. As a result, when applied to non-English repositories, the performance of our approach may decrease. We validate our approach on 199 general English Python projects to ensure its generalizability across various Python projects.
\section{Related Works}
\label{sec:related_work}
In this section, we review the literature on refactoring detection approaches in the commit messages.

\noindent\textbf{Refactoring detection tools.} A variety of tools have been proposed to detect refactoring operations in the history of software~\cite{silva2020refdiff, moghadam2021refdetect, tsantalis2018accurate, tsantalis2018accurate, shiblu2022jsdiffer, atwi2021pyref}. These tools primarily rely on algorithms that match patterns and rules to identify refactoring operations between two versions of the source code. 
Silva~\textit{et al.}~\cite{silva2020refdiff} introduce RefDiff 2.0, a multi-language refactoring detection tool, which supports 10 types of refactoring operations in Java and JavaScript and 9 types of refactoring operations in C languages, achieving an overall precision and recall ranging from 88\% to 91\% across different languages. Moghadam~\textit{et al.}~\cite{moghadam2021refdetect} propose RefDetect, which uses string alignment algorithms to detect 27 types of refactoring in Java and C++. RefDetect achieves an overall precision and recall of 91\% and 85\% in Java and 96\% and 94\% in the C++ language. Tsantalis~\textit{et al.}~\cite{tsantalis2018accurate, tsantalis2020refactoringminer} introduce Rminer, which overcomes coverage constraints by applying pattern matching within the Abstract Syntax Tree (AST) to detect refactoring operations. The current version of Rminer (2.4) can identify up to 99 refactoring operations in the history of Java programs with a precision of 100\% and a recall of 94\%.
Using a method similar to Rminer, Shiblu~\textit{et al.}~\cite{shiblu2022jsdiffer} introduce jsDiffer, which is capable of identifying 18 types of refactoring for JavaScript with a precision of 96\% and a recall of 44\%. Sagar~\textit{et al.}~\cite{sagar2021comparing} use text and code features to detect six types of refactoring operations in the history of Java code and achieve an average precision 81\% and an average recall 75\% in refactoring detection.
Atwi~\textit{et al.}~\cite{atwi2021pyref} use a similar approach to Rminer and introduces PyRef. Despite their performance measures, the primary weakness of tools proposed for languages other than Java is a lack of coverage, meaning they often fail to discover all types of refactorings in the development history. 
In this work, we aim to propose a prototype tool to complement and enhance refactoring detection in the Python language, which is popular in both the ML and non-ML development communities.

\noindent\textbf{Self-admitted refactoring patterns.} AlOmar~\textit{et al.}~\cite{alomar2019can} conduct an experimental study and identify a set of SAR patterns potentially used to describe refactoring commits. They find that commits with SAR patterns make a more significant contribution toward major refactoring activities. Ratzinger~\textit{et al.}~\cite{ratzinger2007mining} identify refactoring commits by investigating keywords like {\it "refactor"} and its extensions, such as {\it "needs refactor"} in the commit messages. Similar to Ratzinger~\textit{et al.}~\cite{ratzinger2007mining}, Stroggylos and Spinellis~\cite{stroggylos2007refactoring} examine commit messages, searching for words stemming from \textit{"refactor"} and label them as refactoring commits.  Zhang \textit{et al.}~\cite{di2018preliminary} distinguish the refactoring commits using 22 keywords adopted from Fowler~\cite{fowler2018refactoring} refactoring categories. AlOmar~\textit{et al.}~\cite{alomar2021toward} investigate self-affirmed refactoring commits by leveraging the keywords obtained from their previous work~\cite{alomar2019can} to investigate self-affirmed refactoring commits in Java projects. Kim~\textit{et al.} \cite{kim2014empirical} conduct a survey study, asking developers what keywords they use to indicate refactoring in their commits. They then utilize 10 keywords to detect refactoring commits from version history and correlate them with code features.
Different from the aforementioned approaches, we employ machine learning techniques to detect refactoring activities within the commit history. We compare our proposed approach with the previous keyword-based methods for detecting refactoring or self-admitted refactoring commits. We demonstrate how keyword selection and pattern matching in commit messages cannot detect all possible refactoring commits.

\noindent\textbf{Refactoring detection in python code.} 
Refactoring detection tools for Python are limited, and to the best of our knowledge, code conversion~\cite{dilhara2022discovering} and AST pattern matching~\cite{atwi2021pyref} are the only approaches available. Due to the differences in syntax between Python and Java, certain code information may be lost, such as the lack of variable type definition in Python. Therefore, code conversion may not be a practical solution. 
Dilhara \textit{et al.}~\cite{dilhara2022discovering} convert Python code into Java and then use Rminer \cite{tsantalis2020refactoringminer} to detect the refactoring operations in Python projects. 
Atwi~\textit{et al.}~\cite{atwi2021pyref} use the same pattern matching in Abstract Syntax Tree (AST) as Rminer \cite{tsantalis2020refactoringminer} does and introduce PyRef, which can identify up to 11 types of refactorings, achieving an overall precision of 89.6\% and a recall of 76.1\%. They examine Dilhara~\textit{et al.}'s approach for converting Python code to Java and observe a precision of 71.74\% and a recall of 50.13\% for their approach. PyRef is the superior refactoring detection tool in Python showing higher precision and recall compared to its opponent. In this study, we compare our refactoring detection approach with PyRef and demonstrate a 49\% higher recall in our model compared to PyRef.



\label{sec:related_works}

\section{Conclusion}
\label{sec:conclusion}

In this study, we introduce our approach, MLRefScanner, which is capable of identifying refactoring commits in the history of ML projects. We conduct extensive experiments on 199 ML Python projects and verify the performance of MLRefScanner in various testing scenarios. MLRefScanner achieves a precision of 94\%, a recall of 82\%, and an AUC of 89\%, and a minimum set of features is provided for distinguishing refactoring commits. Our approach outperforms state-of-the-art refactoring detection methods, with an average of 21\% higher precision, 52\% higher recall, 28\% higher AUC, and 45\% higher F1 scores \fixed{in the mixed-projects scenario.}  Moreover, ensembling MLRefScanner with PyRef, the most recent and accurate refactoring detection tool, showcases a 60\% improvement in its overall recall. MLRefScanner successfully identifies refactoring activities that were previously overlooked by state-of-the-art methods and can detect commits involving ML-specific refactoring operations. MLRefScanner enhances the refactoring detection in ML technical domain by addressing the issue of low recall present in existing state-of-the-art approaches. Moreover, we contribute to the refactoring research community with an extensive dataset that includes refactoring activities from 199 ML repositories for further refactoring analysis.

In the future, we plan to extend and fine-tune our approach to enable the detection of specific types of refactorings, such as pull-up method and move-class, by mining software development histories.


\begin{acks}
We want to express our appreciation to Jonathan Cordeiro at Queen's University for helping to manually validate our results. His commitment was really helpful in confirming the reliability and precision of our findings.
\end{acks}

\bibliographystyle{ACM-Reference-Format}
\bibliography{sample-base}


\begin{thebibliography}{71}


\ifx \showCODEN    \undefined \def \showCODEN     #1{\unskip}     \fi
\ifx \showDOI      \undefined \def \showDOI       #1{#1}\fi
\ifx \showISBNx    \undefined \def \showISBNx     #1{\unskip}     \fi
\ifx \showISBNxiii \undefined \def \showISBNxiii  #1{\unskip}     \fi
\ifx \showISSN     \undefined \def \showISSN      #1{\unskip}     \fi
\ifx \showLCCN     \undefined \def \showLCCN      #1{\unskip}     \fi
\ifx \shownote     \undefined \def \shownote      #1{#1}          \fi
\ifx \showarticletitle \undefined \def \showarticletitle #1{#1}   \fi
\ifx \showURL      \undefined \def \showURL       {\relax}        \fi
\providecommand\bibfield[2]{#2}
\providecommand\bibinfo[2]{#2}
\providecommand\natexlab[1]{#1}
\providecommand\showeprint[2][]{arXiv:#2}

\bibitem[Und(2022)]%
        {Understand}
 \bibinfo{year}{2022}\natexlab{}.
\newblock
\newblock
\urldef\tempurl%
\url{https://support.scitools.com/support/solutions/articles/70000582223-what-metrics-does-understand-have-}
\showURL{%
\tempurl}


\bibitem[TIO(2023)]%
        {TIOBE_2022}
 \bibinfo{year}{2023}\natexlab{}.
\newblock
\newblock
\urldef\tempurl%
\url{https://www.tiobe.com/tiobe-index/}
\showURL{%
\tempurl}


\bibitem[Pyt(2023)]%
        {Python.org}
 \bibinfo{year}{2023}\natexlab{}.
\newblock
\newblock
\urldef\tempurl%
\url{https://www.python.org/doc/essays/blurb/}
\showURL{%
\tempurl}


\bibitem[Abdulkareem and Abboud(2021)]%
        {abdulkareem2021evaluating}
\bibfield{author}{\bibinfo{person}{Sabah~A Abdulkareem} {and} \bibinfo{person}{Ali~J Abboud}.} \bibinfo{year}{2021}\natexlab{}.
\newblock \showarticletitle{Evaluating python, c++, javascript and java programming languages based on software complexity calculator (halstead metrics)}. In \bibinfo{booktitle}{\emph{IOP Conference Series: Materials Science and Engineering}}, Vol.~\bibinfo{volume}{1076}. IOP Publishing, \bibinfo{pages}{012046}.
\newblock


\bibitem[Al~Dallal and Abdin(2017)]%
        {al2017empirical}
\bibfield{author}{\bibinfo{person}{Jehad Al~Dallal} {and} \bibinfo{person}{Anas Abdin}.} \bibinfo{year}{2017}\natexlab{}.
\newblock \showarticletitle{Empirical evaluation of the impact of object-oriented code refactoring on quality attributes: A systematic literature review}.
\newblock \bibinfo{journal}{\emph{IEEE Transactions on Software Engineering}} \bibinfo{volume}{44}, \bibinfo{number}{1} (\bibinfo{year}{2017}), \bibinfo{pages}{44--69}.
\newblock


\bibitem[AlOmar et~al\mbox{.}(2019)]%
        {alomar2019can}
\bibfield{author}{\bibinfo{person}{Eman AlOmar}, \bibinfo{person}{Mohamed~Wiem Mkaouer}, {and} \bibinfo{person}{Ali Ouni}.} \bibinfo{year}{2019}\natexlab{}.
\newblock \showarticletitle{Can refactoring be self-affirmed? an exploratory study on how developers document their refactoring activities in commit messages}. In \bibinfo{booktitle}{\emph{2019 IEEE/ACM 3rd International Workshop on Refactoring (IWoR)}}. IEEE, \bibinfo{pages}{51--58}.
\newblock


\bibitem[AlOmar et~al\mbox{.}(2021a)]%
        {alomar2021refactoring}
\bibfield{author}{\bibinfo{person}{Eman~Abdullah AlOmar}, \bibinfo{person}{Hussein AlRubaye}, \bibinfo{person}{Mohamed~Wiem Mkaouer}, \bibinfo{person}{Ali Ouni}, {and} \bibinfo{person}{Marouane Kessentini}.} \bibinfo{year}{2021}\natexlab{a}.
\newblock \showarticletitle{Refactoring practices in the context of modern code review: An industrial case study at Xerox}. In \bibinfo{booktitle}{\emph{2021 IEEE/ACM 43rd International Conference on Software Engineering: Software Engineering in Practice (ICSE-SEIP)}}. IEEE, \bibinfo{pages}{348--357}.
\newblock


\bibitem[AlOmar et~al\mbox{.}(2021b)]%
        {alomar2021toward}
\bibfield{author}{\bibinfo{person}{Eman~Abdullah AlOmar}, \bibinfo{person}{Mohamed~Wiem Mkaouer}, {and} \bibinfo{person}{Ali Ouni}.} \bibinfo{year}{2021}\natexlab{b}.
\newblock \showarticletitle{Toward the automatic classification of self-affirmed refactoring}.
\newblock \bibinfo{journal}{\emph{Journal of Systems and Software}}  \bibinfo{volume}{171} (\bibinfo{year}{2021}), \bibinfo{pages}{110821}.
\newblock


\bibitem[Amershi et~al\mbox{.}(2019)]%
        {amershi2019software}
\bibfield{author}{\bibinfo{person}{Saleema Amershi}, \bibinfo{person}{Andrew Begel}, \bibinfo{person}{Christian Bird}, \bibinfo{person}{Robert DeLine}, \bibinfo{person}{Harald Gall}, \bibinfo{person}{Ece Kamar}, \bibinfo{person}{Nachiappan Nagappan}, \bibinfo{person}{Besmira Nushi}, {and} \bibinfo{person}{Thomas Zimmermann}.} \bibinfo{year}{2019}\natexlab{}.
\newblock \showarticletitle{Software engineering for machine learning: A case study}. In \bibinfo{booktitle}{\emph{2019 IEEE/ACM 41st International Conference on Software Engineering: Software Engineering in Practice (ICSE-SEIP)}}. IEEE, \bibinfo{pages}{291--300}.
\newblock


\bibitem[Arif and Rana(2020)]%
        {arif2020refactoring}
\bibfield{author}{\bibinfo{person}{Arooj Arif} {and} \bibinfo{person}{Zeeshan~Ali Rana}.} \bibinfo{year}{2020}\natexlab{}.
\newblock \showarticletitle{Refactoring of code to remove technical debt and reduce maintenance effort}. In \bibinfo{booktitle}{\emph{2020 14th International Conference on Open Source Systems and Technologies (ICOSST)}}. IEEE, \bibinfo{pages}{1--7}.
\newblock


\bibitem[Armijo and de~Camargo(2022)]%
        {armijo2022refactoring}
\bibfield{author}{\bibinfo{person}{Guisella~A Armijo} {and} \bibinfo{person}{Valter~V de Camargo}.} \bibinfo{year}{2022}\natexlab{}.
\newblock \showarticletitle{Refactoring Recommendations with Machine Learning}. In \bibinfo{booktitle}{\emph{Anais Estendidos do XXI Simp{\'o}sio Brasileiro de Qualidade de Software}}. SBC, \bibinfo{pages}{15--22}.
\newblock


\bibitem[Atwi et~al\mbox{.}(2021)]%
        {atwi2021pyref}
\bibfield{author}{\bibinfo{person}{Hassan Atwi}, \bibinfo{person}{Bin Lin}, \bibinfo{person}{Nikolaos Tsantalis}, \bibinfo{person}{Yutaro Kashiwa}, \bibinfo{person}{Yasutaka Kamei}, \bibinfo{person}{Naoyasu Ubayashi}, \bibinfo{person}{Gabriele Bavota}, {and} \bibinfo{person}{Michele Lanza}.} \bibinfo{year}{2021}\natexlab{}.
\newblock \showarticletitle{PyRef: refactoring detection in Python projects}. In \bibinfo{booktitle}{\emph{2021 IEEE 21st international working conference on source code analysis and manipulation (SCAM)}}. IEEE, \bibinfo{pages}{136--141}.
\newblock


\bibitem[Balakrishnan and Lloyd-Yemoh(2014)]%
        {balakrishnan2014stemming}
\bibfield{author}{\bibinfo{person}{Vimala Balakrishnan} {and} \bibinfo{person}{Ethel Lloyd-Yemoh}.} \bibinfo{year}{2014}\natexlab{}.
\newblock \showarticletitle{Stemming and lemmatization: A comparison of retrieval performances}.
\newblock  (\bibinfo{year}{2014}).
\newblock


\bibitem[Batista et~al\mbox{.}(2004)]%
        {batista2004study}
\bibfield{author}{\bibinfo{person}{Gustavo~EAPA Batista}, \bibinfo{person}{Ronaldo~C Prati}, {and} \bibinfo{person}{Maria~Carolina Monard}.} \bibinfo{year}{2004}\natexlab{}.
\newblock \showarticletitle{A study of the behavior of several methods for balancing machine learning training data}.
\newblock \bibinfo{journal}{\emph{ACM SIGKDD explorations newsletter}} \bibinfo{volume}{6}, \bibinfo{number}{1} (\bibinfo{year}{2004}), \bibinfo{pages}{20--29}.
\newblock


\bibitem[Bergstra and Bengio(2012)]%
        {bergstra2012random}
\bibfield{author}{\bibinfo{person}{James Bergstra} {and} \bibinfo{person}{Yoshua Bengio}.} \bibinfo{year}{2012}\natexlab{}.
\newblock \showarticletitle{Random search for hyper-parameter optimization.}
\newblock \bibinfo{journal}{\emph{Journal of machine learning research}} \bibinfo{volume}{13}, \bibinfo{number}{2} (\bibinfo{year}{2012}).
\newblock


\bibitem[Cedrim et~al\mbox{.}(2017)]%
        {cedrim2017understanding}
\bibfield{author}{\bibinfo{person}{Diego Cedrim}, \bibinfo{person}{Alessandro Garcia}, \bibinfo{person}{Melina Mongiovi}, \bibinfo{person}{Rohit Gheyi}, \bibinfo{person}{Leonardo Sousa}, \bibinfo{person}{Rafael De~Mello}, \bibinfo{person}{Baldoino Fonseca}, \bibinfo{person}{M{\'a}rcio Ribeiro}, {and} \bibinfo{person}{Alexander Ch{\'a}vez}.} \bibinfo{year}{2017}\natexlab{}.
\newblock \showarticletitle{Understanding the impact of refactoring on smells: A longitudinal study of 23 software projects}. In \bibinfo{booktitle}{\emph{Proceedings of the 2017 11th Joint Meeting on foundations of Software Engineering}}. \bibinfo{pages}{465--475}.
\newblock


\bibitem[Cohen(1988)]%
        {cohen_statistical_1988}
\bibfield{author}{\bibinfo{person}{Jacob Cohen}.} \bibinfo{year}{1988}\natexlab{}.
\newblock \bibinfo{booktitle}{\emph{Statistical {Power} {Analysis} for the {Behavioral} {Sciences}} (\bibinfo{edition}{2} ed.)}.
\newblock \bibinfo{publisher}{Routledge}, \bibinfo{address}{New York}.
\newblock
\showISBNx{978-0-203-77158-7}
\urldef\tempurl%
\url{https://doi.org/10.4324/9780203771587}
\showDOI{\tempurl}


\bibitem[de~Souza~Nascimento et~al\mbox{.}(2019)]%
        {de2019understanding}
\bibfield{author}{\bibinfo{person}{Elizamary de Souza~Nascimento}, \bibinfo{person}{Iftekhar Ahmed}, \bibinfo{person}{Edson Oliveira}, \bibinfo{person}{M{\'a}rcio~Piedade Palheta}, \bibinfo{person}{Igor Steinmacher}, {and} \bibinfo{person}{Tayana Conte}.} \bibinfo{year}{2019}\natexlab{}.
\newblock \showarticletitle{Understanding development process of machine learning systems: Challenges and solutions}. In \bibinfo{booktitle}{\emph{2019 acm/ieee international symposium on empirical software engineering and measurement (esem)}}. IEEE, \bibinfo{pages}{1--6}.
\newblock


\bibitem[Devlin et~al\mbox{.}(2018)]%
        {devlin2018bert}
\bibfield{author}{\bibinfo{person}{Jacob Devlin}, \bibinfo{person}{Ming-Wei Chang}, \bibinfo{person}{Kenton Lee}, {and} \bibinfo{person}{Kristina Toutanova}.} \bibinfo{year}{2018}\natexlab{}.
\newblock \showarticletitle{Bert: Pre-training of deep bidirectional transformers for language understanding}.
\newblock \bibinfo{journal}{\emph{arXiv preprint arXiv:1810.04805}} (\bibinfo{year}{2018}).
\newblock


\bibitem[Di et~al\mbox{.}(2018)]%
        {di2018preliminary}
\bibfield{author}{\bibinfo{person}{Zhang Di}, \bibinfo{person}{Bing Li}, \bibinfo{person}{Zengyang Li}, {and} \bibinfo{person}{Peng Liang}.} \bibinfo{year}{2018}\natexlab{}.
\newblock \showarticletitle{A preliminary investigation of self-admitted refactorings in open source software (S)}. In \bibinfo{booktitle}{\emph{International Conferences on Software Engineering and Knowledge Engineering}}, Vol.~\bibinfo{volume}{2018}. KSI Research Inc. and Knowledge Systems Institute Graduate School, \bibinfo{pages}{165--168}.
\newblock


\bibitem[Dilhara et~al\mbox{.}(2021)]%
        {dilhara2021understanding}
\bibfield{author}{\bibinfo{person}{Malinda Dilhara}, \bibinfo{person}{Ameya Ketkar}, {and} \bibinfo{person}{Danny Dig}.} \bibinfo{year}{2021}\natexlab{}.
\newblock \showarticletitle{Understanding Software-2.0: A Study of Machine Learning library usage and evolution}.
\newblock \bibinfo{journal}{\emph{ACM Transactions on Software Engineering and Methodology (TOSEM)}} \bibinfo{volume}{30}, \bibinfo{number}{4} (\bibinfo{year}{2021}), \bibinfo{pages}{1--42}.
\newblock


\bibitem[Dilhara et~al\mbox{.}(2022)]%
        {dilhara2022discovering}
\bibfield{author}{\bibinfo{person}{Malinda Dilhara}, \bibinfo{person}{Ameya Ketkar}, \bibinfo{person}{Nikhith Sannidhi}, {and} \bibinfo{person}{Danny Dig}.} \bibinfo{year}{2022}\natexlab{}.
\newblock \showarticletitle{Discovering repetitive code changes in Python ML systems}. In \bibinfo{booktitle}{\emph{Proceedings of the 44th International Conference on Software Engineering}}. \bibinfo{pages}{736--748}.
\newblock


\bibitem[Dokmanic et~al\mbox{.}(2015)]%
        {dokmanic2015euclidean}
\bibfield{author}{\bibinfo{person}{Ivan Dokmanic}, \bibinfo{person}{Reza Parhizkar}, \bibinfo{person}{Juri Ranieri}, {and} \bibinfo{person}{Martin Vetterli}.} \bibinfo{year}{2015}\natexlab{}.
\newblock \showarticletitle{Euclidean distance matrices: essential theory, algorithms, and applications}.
\newblock \bibinfo{journal}{\emph{IEEE Signal Processing Magazine}} \bibinfo{volume}{32}, \bibinfo{number}{6} (\bibinfo{year}{2015}), \bibinfo{pages}{12--30}.
\newblock


\bibitem[Drummond et~al\mbox{.}(2003)]%
        {drummond2003c4}
\bibfield{author}{\bibinfo{person}{Chris Drummond}, \bibinfo{person}{Robert~C Holte}, {et~al\mbox{.}}} \bibinfo{year}{2003}\natexlab{}.
\newblock \showarticletitle{C4. 5, class imbalance, and cost sensitivity: why under-sampling beats over-sampling}. In \bibinfo{booktitle}{\emph{Workshop on learning from imbalanced datasets II}}, Vol.~\bibinfo{volume}{11}. \bibinfo{pages}{1--8}.
\newblock


\bibitem[Flesch(1948)]%
        {flesch1948new}
\bibfield{author}{\bibinfo{person}{Rudolph Flesch}.} \bibinfo{year}{1948}\natexlab{}.
\newblock \showarticletitle{A new readability yardstick.}
\newblock \bibinfo{journal}{\emph{Journal of applied psychology}} \bibinfo{volume}{32}, \bibinfo{number}{3} (\bibinfo{year}{1948}), \bibinfo{pages}{221}.
\newblock


\bibitem[Fonseca et~al\mbox{.}(2021)]%
        {fonseca2021addressing}
\bibfield{author}{\bibinfo{person}{Alexandre~Babilone Fonseca}, \bibinfo{person}{David~Correa Martins-Jr}, \bibinfo{person}{Zofia Wicik}, \bibinfo{person}{Marek Postula}, {and} \bibinfo{person}{S{\'e}rgio~Nery Sim{\~o}es}.} \bibinfo{year}{2021}\natexlab{}.
\newblock \showarticletitle{Addressing Classification on Highly Imbalanced Clinical Datasets}. In \bibinfo{booktitle}{\emph{International Conference on Computational Advances in Bio and Medical Sciences}}. Springer, \bibinfo{pages}{103--114}.
\newblock


\bibitem[Fowler(2018)]%
        {fowler2018refactoring}
\bibfield{author}{\bibinfo{person}{Martin Fowler}.} \bibinfo{year}{2018}\natexlab{}.
\newblock \bibinfo{booktitle}{\emph{Refactoring}}.
\newblock \bibinfo{publisher}{Addison-Wesley Professional}.
\newblock


\bibitem[Fowler et~al\mbox{.}(1999)]%
        {fowler1999refactoring}
\bibfield{author}{\bibinfo{person}{Martin Fowler}, \bibinfo{person}{K Beck}, \bibinfo{person}{J Brant}, \bibinfo{person}{W Opdyke}, {and} \bibinfo{person}{D Roberts}.} \bibinfo{year}{1999}\natexlab{}.
\newblock \bibinfo{title}{Refactoring: improving the design of existing code. addison}.
\newblock
\newblock


\bibitem[Gevorkyan et~al\mbox{.}(2019)]%
        {gevorkyan2019review}
\bibfield{author}{\bibinfo{person}{Migran~N Gevorkyan}, \bibinfo{person}{Anastasia~V Demidova}, \bibinfo{person}{Tatiana~S Demidova}, {and} \bibinfo{person}{Anton~A Sobolev}.} \bibinfo{year}{2019}\natexlab{}.
\newblock \showarticletitle{Review and comparative analysis of machine learning libraries for machine learning}.
\newblock \bibinfo{journal}{\emph{Discrete and Continuous Models and Applied Computational Science}} \bibinfo{volume}{27}, \bibinfo{number}{4} (\bibinfo{year}{2019}), \bibinfo{pages}{305--315}.
\newblock


\bibitem[Ghimire et~al\mbox{.}(2021)]%
        {ghimire2021machine}
\bibfield{author}{\bibinfo{person}{Awishkar Ghimire}, \bibinfo{person}{Avinash~Kumar Jha}, \bibinfo{person}{Surendrabikram Thapa}, \bibinfo{person}{Sushruti Mishra}, {and} \bibinfo{person}{Aryan~Mani Jha}.} \bibinfo{year}{2021}\natexlab{}.
\newblock \showarticletitle{Machine learning approach based on hybrid features for detection of phishing URLs}. In \bibinfo{booktitle}{\emph{2021 11th International Conference on Cloud Computing, Data Science \& Engineering (Confluence)}}. IEEE, \bibinfo{pages}{954--959}.
\newblock


\bibitem[Hamou-Lhadj(2008)]%
        {hamou2008measuring}
\bibfield{author}{\bibinfo{person}{Abdelwahab Hamou-Lhadj}.} \bibinfo{year}{2008}\natexlab{}.
\newblock \showarticletitle{Measuring the complexity of traces using shannon entropy}. In \bibinfo{booktitle}{\emph{Fifth International Conference on Information Technology: New Generations (ITNG 2008)}}. IEEE, \bibinfo{pages}{489--494}.
\newblock


\bibitem[Hand(2012)]%
        {hand2012assessing}
\bibfield{author}{\bibinfo{person}{David~J Hand}.} \bibinfo{year}{2012}\natexlab{}.
\newblock \showarticletitle{Assessing the performance of classification methods}.
\newblock \bibinfo{journal}{\emph{International Statistical Review}} \bibinfo{volume}{80}, \bibinfo{number}{3} (\bibinfo{year}{2012}), \bibinfo{pages}{400--414}.
\newblock


\bibitem[Harrell et~al\mbox{.}(2001)]%
        {harrell2001regression}
\bibfield{author}{\bibinfo{person}{Frank~E Harrell} {et~al\mbox{.}}} \bibinfo{year}{2001}\natexlab{}.
\newblock \bibinfo{booktitle}{\emph{Regression modeling strategies: with applications to linear models, logistic regression, and survival analysis}}. Vol.~\bibinfo{volume}{608}.
\newblock \bibinfo{publisher}{Springer}.
\newblock


\bibitem[Hazewinkel(2001)]%
        {hazewinkel2001minimax}
\bibfield{author}{\bibinfo{person}{Michiel Hazewinkel}.} \bibinfo{year}{2001}\natexlab{}.
\newblock \bibinfo{title}{Minimax principle, Encyclopaedia of mathematics}.
\newblock
\newblock


\bibitem[Jafarzadeh et~al\mbox{.}(2021)]%
        {jafarzadeh2021bagging}
\bibfield{author}{\bibinfo{person}{Hamid Jafarzadeh}, \bibinfo{person}{Masoud Mahdianpari}, \bibinfo{person}{Eric Gill}, \bibinfo{person}{Fariba Mohammadimanesh}, {and} \bibinfo{person}{Saeid Homayouni}.} \bibinfo{year}{2021}\natexlab{}.
\newblock \showarticletitle{Bagging and boosting ensemble classifiers for classification of multispectral, hyperspectral and PolSAR data: a comparative evaluation}.
\newblock \bibinfo{journal}{\emph{Remote Sensing}} \bibinfo{volume}{13}, \bibinfo{number}{21} (\bibinfo{year}{2021}), \bibinfo{pages}{4405}.
\newblock


\bibitem[Jiarpakdee et~al\mbox{.}(2016)]%
        {jiarpakdee2016study}
\bibfield{author}{\bibinfo{person}{Jirayus Jiarpakdee}, \bibinfo{person}{Chakkrit Tantithamthavorn}, \bibinfo{person}{Akinori Ihara}, {and} \bibinfo{person}{Kenichi Matsumoto}.} \bibinfo{year}{2016}\natexlab{}.
\newblock \showarticletitle{A study of redundant metrics in defect prediction datasets}. In \bibinfo{booktitle}{\emph{2016 IEEE International Symposium on Software Reliability Engineering Workshops (ISSREW)}}. IEEE, \bibinfo{pages}{51--52}.
\newblock


\bibitem[Khoirom et~al\mbox{.}(2020)]%
        {khoirom2020comparative}
\bibfield{author}{\bibinfo{person}{Selina Khoirom}, \bibinfo{person}{Moirangthem Sonia}, \bibinfo{person}{Borishphia Laikhuram}, \bibinfo{person}{Jaeson Laishram}, {and} \bibinfo{person}{Tekcham~Davidson Singh}.} \bibinfo{year}{2020}\natexlab{}.
\newblock \showarticletitle{Comparative analysis of Python and Java for beginners}.
\newblock \bibinfo{journal}{\emph{Int. Res. J. Eng. Technol}} \bibinfo{volume}{7}, \bibinfo{number}{8} (\bibinfo{year}{2020}), \bibinfo{pages}{4384--4407}.
\newblock


\bibitem[Kim et~al\mbox{.}(2014)]%
        {kim2014empirical}
\bibfield{author}{\bibinfo{person}{Miryung Kim}, \bibinfo{person}{Thomas Zimmermann}, {and} \bibinfo{person}{Nachiappan Nagappan}.} \bibinfo{year}{2014}\natexlab{}.
\newblock \showarticletitle{An empirical study of refactoringchallenges and benefits at microsoft}.
\newblock \bibinfo{journal}{\emph{IEEE Transactions on Software Engineering}} \bibinfo{volume}{40}, \bibinfo{number}{7} (\bibinfo{year}{2014}), \bibinfo{pages}{633--649}.
\newblock


\bibitem[Kotsiantis et~al\mbox{.}(2006)]%
        {kotsiantis2006handling}
\bibfield{author}{\bibinfo{person}{Sotiris Kotsiantis}, \bibinfo{person}{Dimitris Kanellopoulos}, \bibinfo{person}{Panayiotis Pintelas}, {et~al\mbox{.}}} \bibinfo{year}{2006}\natexlab{}.
\newblock \showarticletitle{Handling imbalanced datasets: A review}.
\newblock \bibinfo{journal}{\emph{GESTS international transactions on computer science and engineering}} \bibinfo{volume}{30}, \bibinfo{number}{1} (\bibinfo{year}{2006}), \bibinfo{pages}{25--36}.
\newblock


\bibitem[Luhn(1957)]%
        {luhn1957statistical}
\bibfield{author}{\bibinfo{person}{Hans~Peter Luhn}.} \bibinfo{year}{1957}\natexlab{}.
\newblock \showarticletitle{A statistical approach to mechanized encoding and searching of literary information}.
\newblock \bibinfo{journal}{\emph{IBM Journal of research and development}} \bibinfo{volume}{1}, \bibinfo{number}{4} (\bibinfo{year}{1957}), \bibinfo{pages}{309--317}.
\newblock


\bibitem[Mani and Zhang(2003)]%
        {mani2003knn}
\bibfield{author}{\bibinfo{person}{Inderjeet Mani} {and} \bibinfo{person}{I Zhang}.} \bibinfo{year}{2003}\natexlab{}.
\newblock \showarticletitle{kNN approach to unbalanced data distributions: a case study involving information extraction}. In \bibinfo{booktitle}{\emph{Proceedings of workshop on learning from imbalanced datasets}}, Vol.~\bibinfo{volume}{126}. ICML, \bibinfo{pages}{1--7}.
\newblock


\bibitem[McHugh(2012)]%
        {mchugh2012interrater}
\bibfield{author}{\bibinfo{person}{Mary~L McHugh}.} \bibinfo{year}{2012}\natexlab{}.
\newblock \showarticletitle{Interrater reliability: the kappa statistic}.
\newblock \bibinfo{journal}{\emph{Biochemia medica}} \bibinfo{volume}{22}, \bibinfo{number}{3} (\bibinfo{year}{2012}), \bibinfo{pages}{276--282}.
\newblock


\bibitem[Medeiros et~al\mbox{.}(2023)]%
        {medeiros2023trustworthiness}
\bibfield{author}{\bibinfo{person}{Nadia Medeiros}, \bibinfo{person}{Naghmeh Ivaki}, \bibinfo{person}{Pedro Costa}, {and} \bibinfo{person}{Marco Vieira}.} \bibinfo{year}{2023}\natexlab{}.
\newblock \showarticletitle{Trustworthiness models to categorize and prioritize code for security improvement}.
\newblock \bibinfo{journal}{\emph{Journal of Systems and Software}}  \bibinfo{volume}{198} (\bibinfo{year}{2023}), \bibinfo{pages}{111621}.
\newblock


\bibitem[Mihajlovi{\'c} et~al\mbox{.}(2020)]%
        {mihajlovic2020use}
\bibfield{author}{\bibinfo{person}{S Mihajlovi{\'c}}, \bibinfo{person}{A Kupusinac}, \bibinfo{person}{D Iveti{\'c}}, {and} \bibinfo{person}{I Berkovi{\'c}}.} \bibinfo{year}{2020}\natexlab{}.
\newblock \showarticletitle{The Use of Python in the field of Artifical Intelligence}. In \bibinfo{booktitle}{\emph{International Conference on Information Technology and Development of Education--ITRO}}.
\newblock


\bibitem[Mikolov et~al\mbox{.}(2013)]%
        {mikolov2013linguistic}
\bibfield{author}{\bibinfo{person}{Tom{\'a}{\v{s}} Mikolov}, \bibinfo{person}{Wen-tau Yih}, {and} \bibinfo{person}{Geoffrey Zweig}.} \bibinfo{year}{2013}\natexlab{}.
\newblock \showarticletitle{Linguistic regularities in continuous space word representations}. In \bibinfo{booktitle}{\emph{Proceedings of the 2013 conference of the north american chapter of the association for computational linguistics: Human language technologies}}. \bibinfo{pages}{746--751}.
\newblock


\bibitem[Miles(2005)]%
        {miles2005r}
\bibfield{author}{\bibinfo{person}{Jeremy Miles}.} \bibinfo{year}{2005}\natexlab{}.
\newblock \showarticletitle{R-squared, adjusted R-squared}.
\newblock \bibinfo{journal}{\emph{Encyclopedia of statistics in behavioral science}} (\bibinfo{year}{2005}).
\newblock


\bibitem[Mishra(2017)]%
        {mishra2017handling}
\bibfield{author}{\bibinfo{person}{Satwik Mishra}.} \bibinfo{year}{2017}\natexlab{}.
\newblock \showarticletitle{Handling imbalanced data: SMOTE vs. random undersampling}.
\newblock \bibinfo{journal}{\emph{Int. Res. J. Eng. Technol}} \bibinfo{volume}{4}, \bibinfo{number}{8} (\bibinfo{year}{2017}), \bibinfo{pages}{317--320}.
\newblock


\bibitem[Moghadam et~al\mbox{.}(2021)]%
        {moghadam2021refdetect}
\bibfield{author}{\bibinfo{person}{Iman~Hemati Moghadam}, \bibinfo{person}{Mel~{\'O} Cinn{\'e}ide}, \bibinfo{person}{Faezeh Zarepour}, {and} \bibinfo{person}{Mohamad~Aref Jahanmir}.} \bibinfo{year}{2021}\natexlab{}.
\newblock \showarticletitle{Refdetect: A multi-language refactoring detection tool based on string alignment}.
\newblock \bibinfo{journal}{\emph{IEEE Access}}  \bibinfo{volume}{9} (\bibinfo{year}{2021}), \bibinfo{pages}{86698--86727}.
\newblock


\bibitem[Mor{\'a}n-Fern{\'a}ndez et~al\mbox{.}(2022)]%
        {moran2022important}
\bibfield{author}{\bibinfo{person}{Laura Mor{\'a}n-Fern{\'a}ndez}, \bibinfo{person}{Ver{\'o}nica B{\'o}lon-Canedo}, {and} \bibinfo{person}{Amparo Alonso-Betanzos}.} \bibinfo{year}{2022}\natexlab{}.
\newblock \showarticletitle{How important is data quality? Best classifiers vs best features}.
\newblock \bibinfo{journal}{\emph{Neurocomputing}}  \bibinfo{volume}{470} (\bibinfo{year}{2022}), \bibinfo{pages}{365--375}.
\newblock


\bibitem[Nguyen et~al\mbox{.}(2010)]%
        {nguyen2010studying}
\bibfield{author}{\bibinfo{person}{Thanh~HD Nguyen}, \bibinfo{person}{Bram Adams}, {and} \bibinfo{person}{Ahmed~E Hassan}.} \bibinfo{year}{2010}\natexlab{}.
\newblock \showarticletitle{Studying the impact of dependency network measures on software quality}. In \bibinfo{booktitle}{\emph{2010 IEEE International Conference on Software Maintenance}}. IEEE, \bibinfo{pages}{1--10}.
\newblock


\bibitem[Noei et~al\mbox{.}(2019)]%
        {noei2019too}
\bibfield{author}{\bibinfo{person}{Ehsan Noei}, \bibinfo{person}{Feng Zhang}, {and} \bibinfo{person}{Ying Zou}.} \bibinfo{year}{2019}\natexlab{}.
\newblock \showarticletitle{Too many user-reviews, what should app developers look at first?}
\newblock \bibinfo{journal}{\emph{IEEE Transactions on Software Engineering}} (\bibinfo{year}{2019}).
\newblock


\bibitem[Nyamawe et~al\mbox{.}(2020)]%
        {nyamawe2020feature}
\bibfield{author}{\bibinfo{person}{Ally~S Nyamawe}, \bibinfo{person}{Hui Liu}, \bibinfo{person}{Nan Niu}, \bibinfo{person}{Qasim Umer}, {and} \bibinfo{person}{Zhendong Niu}.} \bibinfo{year}{2020}\natexlab{}.
\newblock \showarticletitle{Feature requests-based recommendation of software refactorings}.
\newblock \bibinfo{journal}{\emph{Empirical Software Engineering}}  \bibinfo{volume}{25} (\bibinfo{year}{2020}), \bibinfo{pages}{4315--4347}.
\newblock


\bibitem[Openja et~al\mbox{.}(2022)]%
        {openja2022studying}
\bibfield{author}{\bibinfo{person}{Moses Openja}, \bibinfo{person}{Forough Majidi}, \bibinfo{person}{Foutse Khomh}, \bibinfo{person}{Bhagya Chembakottu}, {and} \bibinfo{person}{Heng Li}.} \bibinfo{year}{2022}\natexlab{}.
\newblock \showarticletitle{Studying the practices of deploying machine learning projects on docker}. In \bibinfo{booktitle}{\emph{Proceedings of the 26th International Conference on Evaluation and Assessment in Software Engineering}}. \bibinfo{pages}{190--200}.
\newblock


\bibitem[Pintas et~al\mbox{.}(2021)]%
        {pintas2021feature}
\bibfield{author}{\bibinfo{person}{Julliano~Trindade Pintas}, \bibinfo{person}{Leandro~AF Fernandes}, {and} \bibinfo{person}{Ana Cristina~Bicharra Garcia}.} \bibinfo{year}{2021}\natexlab{}.
\newblock \showarticletitle{Feature selection methods for text classification: a systematic literature review}.
\newblock \bibinfo{journal}{\emph{Artificial Intelligence Review}} \bibinfo{volume}{54}, \bibinfo{number}{8} (\bibinfo{year}{2021}), \bibinfo{pages}{6149--6200}.
\newblock


\bibitem[Raschka et~al\mbox{.}(2020)]%
        {raschka2020machine}
\bibfield{author}{\bibinfo{person}{Sebastian Raschka}, \bibinfo{person}{Joshua Patterson}, {and} \bibinfo{person}{Corey Nolet}.} \bibinfo{year}{2020}\natexlab{}.
\newblock \showarticletitle{Machine learning in python: Main developments and technology trends in data science, machine learning, and artificial intelligence}.
\newblock \bibinfo{journal}{\emph{Information}} \bibinfo{volume}{11}, \bibinfo{number}{4} (\bibinfo{year}{2020}), \bibinfo{pages}{193}.
\newblock


\bibitem[Ratzinger et~al\mbox{.}(2007)]%
        {ratzinger2007mining}
\bibfield{author}{\bibinfo{person}{Jacek Ratzinger}, \bibinfo{person}{Thomas Sigmund}, \bibinfo{person}{Peter Vorburger}, {and} \bibinfo{person}{Harald Gall}.} \bibinfo{year}{2007}\natexlab{}.
\newblock \showarticletitle{Mining software evolution to predict refactoring}. In \bibinfo{booktitle}{\emph{First International Symposium on Empirical Software Engineering and Measurement (ESEM 2007)}}. IEEE, \bibinfo{pages}{354--363}.
\newblock


\bibitem[Ribeiro et~al\mbox{.}(2016)]%
        {ribeiro2016should}
\bibfield{author}{\bibinfo{person}{Marco~Tulio Ribeiro}, \bibinfo{person}{Sameer Singh}, {and} \bibinfo{person}{Carlos Guestrin}.} \bibinfo{year}{2016}\natexlab{}.
\newblock \showarticletitle{" Why should i trust you?" Explaining the predictions of any classifier}. In \bibinfo{booktitle}{\emph{Proceedings of the 22nd ACM SIGKDD international conference on knowledge discovery and data mining}}. \bibinfo{pages}{1135--1144}.
\newblock


\bibitem[Rokach(2010)]%
        {rokach2010ensemble}
\bibfield{author}{\bibinfo{person}{Lior Rokach}.} \bibinfo{year}{2010}\natexlab{}.
\newblock \showarticletitle{Ensemble-based classifiers}.
\newblock \bibinfo{journal}{\emph{Artificial intelligence review}}  \bibinfo{volume}{33} (\bibinfo{year}{2010}), \bibinfo{pages}{1--39}.
\newblock


\bibitem[Sagar et~al\mbox{.}(2021)]%
        {sagar2021comparing}
\bibfield{author}{\bibinfo{person}{Priyadarshni~Suresh Sagar}, \bibinfo{person}{Eman~Abdulah AlOmar}, \bibinfo{person}{Mohamed~Wiem Mkaouer}, \bibinfo{person}{Ali Ouni}, {and} \bibinfo{person}{Christian~D Newman}.} \bibinfo{year}{2021}\natexlab{}.
\newblock \showarticletitle{Comparing commit messages and source code metrics for the prediction refactoring activities}.
\newblock \bibinfo{journal}{\emph{Algorithms}} \bibinfo{volume}{14}, \bibinfo{number}{10} (\bibinfo{year}{2021}), \bibinfo{pages}{289}.
\newblock


\bibitem[Shannon(1948)]%
        {shannon1948mathematical}
\bibfield{author}{\bibinfo{person}{Claude~Elwood Shannon}.} \bibinfo{year}{1948}\natexlab{}.
\newblock \showarticletitle{A mathematical theory of communication}.
\newblock \bibinfo{journal}{\emph{The Bell system technical journal}} \bibinfo{volume}{27}, \bibinfo{number}{3} (\bibinfo{year}{1948}), \bibinfo{pages}{379--423}.
\newblock


\bibitem[Shiblu(2022)]%
        {shiblu2022jsdiffer}
\bibfield{author}{\bibinfo{person}{Mosabbir~Khan Shiblu}.} \bibinfo{year}{2022}\natexlab{}.
\newblock \emph{\bibinfo{title}{JsDiffer: Refactoring Detection in JavaScript}}.
\newblock \bibinfo{thesistype}{Ph.\,D. Dissertation}. \bibinfo{school}{Concordia University Montr{\'e}al, Qu{\'e}bec, Canada}.
\newblock


\bibitem[Silva et~al\mbox{.}(2020)]%
        {silva2020refdiff}
\bibfield{author}{\bibinfo{person}{Danilo Silva}, \bibinfo{person}{Joao~Paulo da Silva}, \bibinfo{person}{Gustavo Santos}, \bibinfo{person}{Ricardo Terra}, {and} \bibinfo{person}{Marco~Tulio Valente}.} \bibinfo{year}{2020}\natexlab{}.
\newblock \showarticletitle{Refdiff 2.0: A multi-language refactoring detection tool}.
\newblock \bibinfo{journal}{\emph{IEEE Transactions on Software Engineering}} \bibinfo{volume}{47}, \bibinfo{number}{12} (\bibinfo{year}{2020}), \bibinfo{pages}{2786--2802}.
\newblock


\bibitem[Sparck~Jones(1972)]%
        {sparck1972statistical}
\bibfield{author}{\bibinfo{person}{Karen Sparck~Jones}.} \bibinfo{year}{1972}\natexlab{}.
\newblock \showarticletitle{A statistical interpretation of term specificity and its application in retrieval}.
\newblock \bibinfo{journal}{\emph{Journal of documentation}} \bibinfo{volume}{28}, \bibinfo{number}{1} (\bibinfo{year}{1972}), \bibinfo{pages}{11--21}.
\newblock


\bibitem[Srinath(2017)]%
        {srinath2017python}
\bibfield{author}{\bibinfo{person}{KR Srinath}.} \bibinfo{year}{2017}\natexlab{}.
\newblock \showarticletitle{Python--the fastest growing programming language}.
\newblock \bibinfo{journal}{\emph{International Research Journal of Engineering and Technology}} \bibinfo{volume}{4}, \bibinfo{number}{12} (\bibinfo{year}{2017}), \bibinfo{pages}{354--357}.
\newblock


\bibitem[Stroggylos and Spinellis(2007)]%
        {stroggylos2007refactoring}
\bibfield{author}{\bibinfo{person}{Konstantinos Stroggylos} {and} \bibinfo{person}{Diomidis Spinellis}.} \bibinfo{year}{2007}\natexlab{}.
\newblock \showarticletitle{Refactoring--does it improve software quality?}. In \bibinfo{booktitle}{\emph{Fifth International Workshop on Software Quality (WoSQ'07: ICSE Workshops 2007)}}. IEEE, \bibinfo{pages}{10--10}.
\newblock


\bibitem[Thomas({[n.\,d.]})]%
        {Thomas}
\bibfield{author}{\bibinfo{person}{Reuben Thomas}.} \bibinfo{year}{[n.\,d.]}\natexlab{}.
\newblock
\newblock
\urldef\tempurl%
\url{https://abiword.github.io/enchant/}
\showURL{%
\tempurl}


\bibitem[Tsantalis et~al\mbox{.}(2020)]%
        {tsantalis2020refactoringminer}
\bibfield{author}{\bibinfo{person}{Nikolaos Tsantalis}, \bibinfo{person}{Ameya Ketkar}, {and} \bibinfo{person}{Danny Dig}.} \bibinfo{year}{2020}\natexlab{}.
\newblock \showarticletitle{RefactoringMiner 2.0}.
\newblock \bibinfo{journal}{\emph{IEEE Transactions on Software Engineering}} \bibinfo{volume}{48}, \bibinfo{number}{3} (\bibinfo{year}{2020}), \bibinfo{pages}{930--950}.
\newblock


\bibitem[Tsantalis et~al\mbox{.}(2018)]%
        {tsantalis2018accurate}
\bibfield{author}{\bibinfo{person}{Nikolaos Tsantalis}, \bibinfo{person}{Matin Mansouri}, \bibinfo{person}{Laleh~M Eshkevari}, \bibinfo{person}{Davood Mazinanian}, {and} \bibinfo{person}{Danny Dig}.} \bibinfo{year}{2018}\natexlab{}.
\newblock \showarticletitle{Accurate and efficient refactoring detection in commit history}. In \bibinfo{booktitle}{\emph{Proceedings of the 40th international conference on software engineering}}. \bibinfo{pages}{483--494}.
\newblock


\bibitem[Vatanapakorn et~al\mbox{.}(2022)]%
        {vatanapakorn2022python}
\bibfield{author}{\bibinfo{person}{Natthida Vatanapakorn}, \bibinfo{person}{Chitsutha Soomlek}, {and} \bibinfo{person}{Pusadee Seresangtakul}.} \bibinfo{year}{2022}\natexlab{}.
\newblock \showarticletitle{Python Code Smell Detection Using Machine Learning}. In \bibinfo{booktitle}{\emph{2022 26th International Computer Science and Engineering Conference (ICSEC)}}. IEEE, \bibinfo{pages}{128--133}.
\newblock


\bibitem[Wan et~al\mbox{.}(2019)]%
        {wan2019does}
\bibfield{author}{\bibinfo{person}{Zhiyuan Wan}, \bibinfo{person}{Xin Xia}, \bibinfo{person}{David Lo}, {and} \bibinfo{person}{Gail~C Murphy}.} \bibinfo{year}{2019}\natexlab{}.
\newblock \showarticletitle{How does machine learning change software development practices?}
\newblock \bibinfo{journal}{\emph{IEEE Transactions on Software Engineering}} \bibinfo{volume}{47}, \bibinfo{number}{9} (\bibinfo{year}{2019}), \bibinfo{pages}{1857--1871}.
\newblock


\bibitem[Zhou et~al\mbox{.}(2021)]%
        {zhou2021investigating}
\bibfield{author}{\bibinfo{person}{Andy Zhou}, \bibinfo{person}{Kazi~Zakia Sultana}, {and} \bibinfo{person}{Bharath~K Samanthula}.} \bibinfo{year}{2021}\natexlab{}.
\newblock \showarticletitle{Investigating the changes in software metrics after vulnerability is fixed}. In \bibinfo{booktitle}{\emph{2021 IEEE International Conference on Big Data (Big Data)}}. IEEE, \bibinfo{pages}{5658--5663}.
\newblock


\end{thebibliography}

\end{document}